# Topological insulators on complex networks

Sunkyu Yu[1†], Xianji Piao[2§], and Namkyoo Park[3*]

[1]Intelligent Wave Systems Laboratory, Department of Electrical and Computer Engineering, Seoul National University, Seoul 08826, Korea

[2]Wave Engineering Laboratory, School of Electrical and Computer Engineering, University of Seoul, Seoul 02504, Korea

[3]Photonic Systems Laboratory, Department of Electrical and Computer Engineering, Seoul National University, Seoul 08826, Korea

E-mail address for correspondence: sunkyu.yu@snu.ac.kr; piao@uos.ac.kr; nkpark@snu.ac.kr

**Abstract**

The landscape of topological insulators has expanded beyond its traditional domain of periodic lattices with short-range hopping. Related studies have explored topological phenomena under non-Euclidean geometries, disordered structures, and long-range hopping, progressively narrowing the gap between topological phases of matter and complex networks. Here we demonstrate that genuine complex networks—far beyond periodic lattices and their conventional variants—can themselves host topological insulating phases. By developing a nonconflicting design framework for multiple topological objectives, we realize topological insulators in high-degree Hofstadter lattices and across regular, small-world, and random networks. This generalization uncovers distinct network-specific features, including an expanded range of attainable topological characteristics, a degree-dependent transition from Hofstadter-type to Haldane-type Chern insulators, and, most notably, the superior robustness and reconfigurability of small-world topological insulators. Our graph-theoretic framework establishes network complexity as a resource for realizing high-capacity and reconfigurable topological insulators.

## Introduction

Recent efforts to generalize topological insulators (TIs) have pushed their domain of realization beyond canonical models[1]—including the Hofstadter[2], Haldane[3], and Kane-Mele[4] models—which are formulated on periodic lattices with predominantly short-range hopping. To encompass diverse generalizations in this field, a network formulation—mapping sites onto nodes and hopping amplitudes onto links—provides a natural language for their unified classification. For example, topological hyperbolic lattices expand the accessible range of node degrees in two-dimensional (2D) regular networks[5-8]. Network irregularity encompasses both local defects, such as disclinations[9] and dislocations[10], and statistically homogeneous disorder, whether uncorrelated[11-13] or correlated[14], in TIs. Beyond the regime of short-range links, nonlocal hopping, which corresponds to network shortcuts, has been explored to realize flatbands[15], large Chern numbers[16,17], and corner states[18]. To investigate such platforms, considerable effort has been devoted to real-space topological diagnostics[19-22].

These achievements, gradually bridging the gap between topological physics[1] and network science[23], motivate exploring topologically protected signal transport in generic networks—structures characterized by a fundamentally different notion of topology: network topology[23]. Although topologically protected transport may offer a promising route to defect-immune information flow resilient to network-level complexity, severe challenges arise, even at the theoretical level, from the very nature of networks. Because a generic network is an abstract object with no intrinsic embedding in real space, the dimensional classification of topological phases[24,25] becomes ill-defined, thereby hindering the direct application of conventional diagnostics[1,19-22]. This obstacle has confined most prior research to highly restricted classes of networks—those with narrowly distributed node degrees and predominantly local connectivity, occasionally supplemented by sparse long-range hopping—in which the Hamiltonian inherits an effective dimensionality and locality. Although topological phases in heavy-tailed networks[26] have recently been identified[27], topologically protected transport has yet to be demonstrated even in this highly symmetric setting. Consequently, whether topologically protected transport persists across the broader spectrum of network topologies remains an open question.

Here, we demonstrate the existence of topologically protected signal transport in complex networks, inheriting the fundamental physics of 2D TIs. We develop a design framework that addresses two major hurdles in exploring topological phases of complex networks: quantized topological invariants and their conflict with robustness. We apply this framework first to the high-degree Hofstadter model, in which nonlocality serves as a resource for extended transport channels and enhanced robustness. More generally, we construct complex networks spanning regular, small-world, and random architectures to support topologically protected directional transport, thereby revealing distinct network-specific phenomena: the critical role of node-degree-enhanced local gaps, a node-degree-driven phase transition in Chern insulators, and the selective sensitivity coupled with enhanced robustness in small-world TIs. These results pave the way for generalized topological physics in complex systems.

## Results

### Designing network TIs

Our goal is to construct network TIs that support topologically protected directional transport analogous to chiral transport in 2D TIs, despite high node degree $k$ and strong randomness with shortcut links. Figure 1a, previewing the design outcome in Fig. 3, illustrates our proposal: topological heterogeneity in complex networks—with nodes carrying nontrivial (red) and trivial (grey) 2D topologies—distinguishes a topologically nontrivial bulk from edge nodes, driving the emergence of protected directional states along the network edge. To validate this proposal, we investigate an $N$-site magnetic graph[27-30]—a network with U(1) hopping phases, where opposite-direction hopping amplitudes are complex conjugates. In the site basis $\{|n\rangle: n = 1, 2, \ldots, N\}$, the Hamiltonian is

$$H = \sum_n \varepsilon_n |n\rangle\langle n| + \sum_{\langle m,n\rangle} t_{mn} e^{-i\varphi_{mn}} |m\rangle\langle n| + h.c., \tag{1}$$

where $\varepsilon_n$ is the onsite energy, $t_{mn}$ and $\varphi_{mn}$ are the hopping magnitude and phase, respectively, $\langle m,n\rangle$ denotes a pair of connected sites, and *h.c.* denotes the Hermitian conjugate. To explore 2D topological physics, we assign 2D coordinates $(x_n, y_n)$ to the $n$-th site, while preserving the connectivity $\langle m,n\rangle$, allowing long-range hopping and link crossings.

To diagnose topological properties in complex networks, we focus on two topological metrics[21,22] (see Methods):

$$C_{\mathrm{L}}(n,E;\kappa) = \frac{1}{2}\mathrm{sig}\big(L(n,E;\kappa)\big), \quad \mu_{\mathrm{L}}(n,E;\kappa) = \min\Big[\big|\mathrm{Spec}\big(L(n,E;\kappa)\big)\big|\Big], \tag{2}$$

where $L(n, E; \kappa)$ denotes the spectral localizer at $(x_n, y_n, E)$, sig($A$) and Spec($A$) denote the signature and spectrum of $A$, respectively, and $\kappa$ sets the relative scale of the spatial and energetic contributions. While the local Chern index $C_{\mathrm{L}}$ characterizes the position-energy local topology, its robustness is quantified by the local gap $\mu_{\mathrm{L}}$. With the connectivity and node features fixed—$t_{mn}$ = 1 for $\langle m,n\rangle$ and $\varepsilon_n = 0$—we seek an inverse-design framework: tailoring the design variables $\{\tau_p\} = \{\{\varphi_{mn}\}, \{(x_n, y_n)\}\}$ to realize prescribed distributions of $C_{\mathrm{L}}$ and $\mu_{\mathrm{L}}$.

The design of topological systems poses several challenges. Most conventional strategies rely on band-theoretic concepts[1,31,32]: targeting unit-cell geometry[33], extending crystalline designs to disordered systems[14,34], or employing statistical approaches[35,36]. Gauge-field designs reach non-Euclidean[5] and heavy-tailed networks[27] but remain restricted to crossing-free planar graphs. Numerically, because topological invariants are piecewise constant in design space, their gradients vanish or are undefined at transitions, hindering gradient-based optimization. Moreover, topological invariant generally competes with the spectral gap, limiting information capacity[17,37]. To address these challenges, we develop a gradient-based, conflict-free multi-objective inverse-design framework.

To overcome the difficulty posed by quantized $C_L$, we introduce the surrogate, $C_\beta(n, E; \kappa) = (1/2)\mathrm{Tr}[\tanh(\beta L)]$, which recovers the index as $C_L = \lim_{\beta\to\infty} C_\beta$ (Fig. 1b). Combining this surrogate with the Hellmann-Feynman theorem[38], we obtain the parameter-space gradients, $\nabla C_\beta$ and $\nabla\mu_L$ (see Methods). Notably, $\nabla C_\beta$ provides a surrogate design direction that steers the system towards the target transition in $C_L$, particularly for large $\beta$.

The conflict between $C_L$ and $\mu_L$ is resolved by the Jacobian-descent scheme (JDS)[39,40] (Supplementary Note S1), which identifies a parameter-update direction along which both objectives are guaranteed to be nonincreasing to first order. We use the gradient-descent scheme (GDS) to achieve $C_L = C_L^0$, followed by JDS to enhance $\mu_L$ to $\mu_L^0$ (Fig. 1c), thereby improving robustness while at least maintaining $C_L$. The framework yields robust, topologically nontrivial bulk nodes within the network—aligned with our proposal for network TIs.

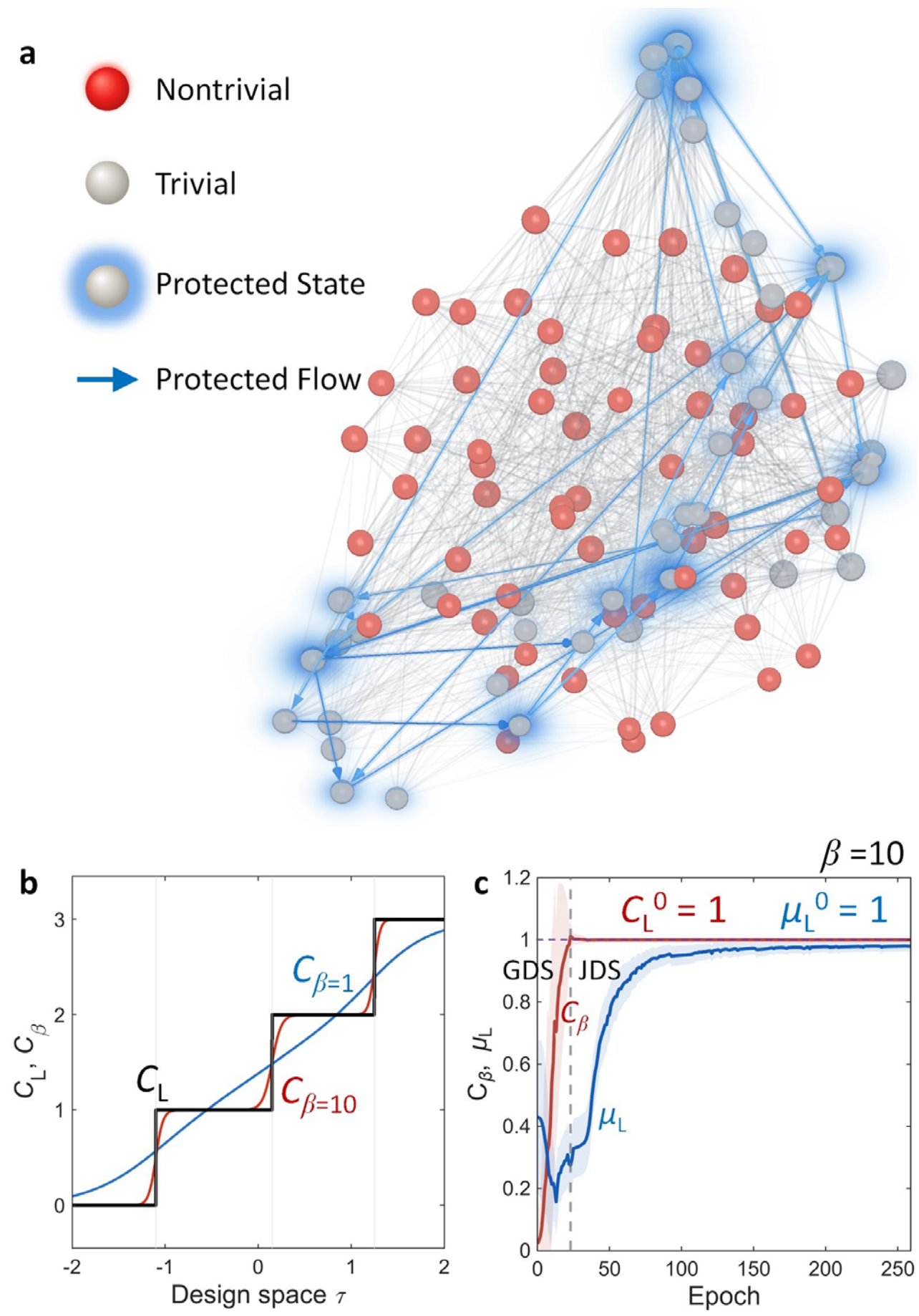

**Fig. 1. Designing network TIs**. **a,** Small-world TI supporting a protected directional state (blue glows and arrows)—the network analogue of chiral edge states in 2D TIs. The realization is one instance of the designs in Fig. 3, for $k = 20$ and link-swap parameter $\rho = 0.1$. **b,** Surrogate model $C_\beta$ for $C_L$ across the design space $\tau$. **c,** Optimization curves for **a**.

## High-degree Hofstadter networks

Using our design framework, we begin with a network extension of a canonical model: the Hofstadter lattice[2]. We use this model to examine how node degree affects topological phenomena by introducing hopping beyond nearest-neighbour (NN) interactions: next-nearest-neighbour (NNN) and the $q$th-nearest-neighbour ($q$NN) hopping (Fig. 2a). In contrast to most previous studies[15,16,41], we assume uniform hopping magnitudes, thereby emulating an ideal abstract network whose links are independent of spatial separation. This construction yields quasi-regular topological networks with varying node degrees: $k = 4$ (NN), 8 (NNN), 12 (3NN), and 20 (4NN).

To establish a baseline, we analyse the standard Hofstadter model using the spectral localizer (Supplementary Note S2). Based on the resulting bulk-edge partition (red dashed box in Fig. 2b), we employ the procedure in Fig. 1c with the design variables $\{\tau_p\} = \{\{\varphi_{mn}\}\}$ to achieve the bulk $C_L^0$, while maximizing the bulk local gap by setting $\mu_L^0 = 1$. To evaluate convergence stability, we initialize the hopping phases using a uniform random distribution. Figure 2b presents a degree-20 (4NN) Hofstadter network, realizing $C_L \approx 3$ throughout the bulk. Rescaling the optimized hopping phases by a factor $\chi$ produces a network counterpart of the Hofstadter butterfly, featuring an eagle-eye-shaped topological gap (Fig. 2c and Supplementary Note S3).

We examine the impact of the hopping range across different $C_L^0$, demonstrating that network connectivity operates as a resource for topological design. Figures 2d and 2e compare the ensemble-averaged bulk $C_L$ and the bulk and edge values of $\mu_L$, respectively, of the high-degree models with those of the standard Hofstadter model ('Hof.') (Supplementary Note S4 for different $\beta$). Even for the same NN connectivity, our framework yields larger local gaps than the standard model (Fig. 2e, 'NN' versus 'Hof.'). Longer-range hopping can further increase the local gap (Fig. 2e, 'NNN') and support higher Chern indices (Fig. 2d, '4NN'), enabling a greater number of topological channels with enhanced robustness. Increasing the node degree therefore supports high-capacity topologically protected transport, which we will again confirm in genuine complex networks.

To clarify the operating principle of the optimized networks, we evaluate the gauge-flux distributions over their closed loops (Supplementary Note S5). The NN design retains a Hofstadter-like flux distribution, consistent with its $C_L$ and $\mu_L$ distributions (Supplementary Note

S3). By contrast, as the hopping range increases, the flux distributions exhibit a strongly suppressed net magnetic flux, suggesting that topological networks containing crossing links more closely resemble magnetic-field-free Chern insulators of the Haldane type.

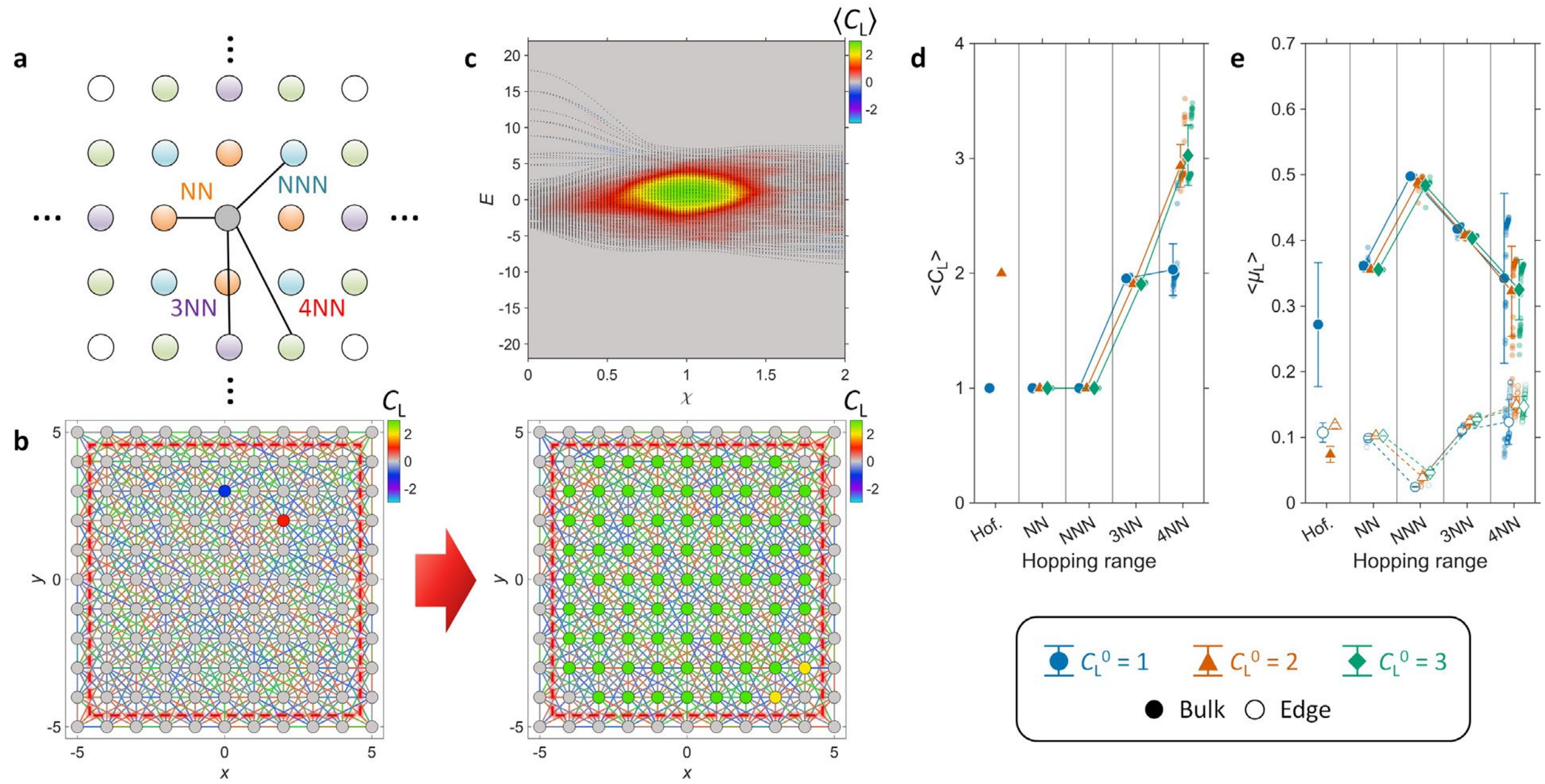

**Fig. 2. High-degree Hofstadter lattices enhance information capacity**. **a,** Hopping shells. **b,** Optimization of a 4NN network toward $C_L{}^0 = 3$. Node and link colours indicate $C_L$ and the directed hopping phase (grey for 0, orange for π/2, green for π, and blue for 3π/2). **c,** Network Hofstadter butterfly. Points denote eigenenergies. $\beta = 1$ in **b** and **c**. **d,e,** Ensemble-averaged bulk $C_L$ (**d**) and $\mu_L$ for bulk (filled) and edge (open) nodes (**e**): 50 realizations per condition. Blue circles, orange triangles, and green diamonds denote $C_L{}^0 = 1$, 2, and 3, respectively. Faint points show individual realizations; error bars, one standard deviation.

## Genuine network TIs

The degree-extended Hofstadter models motivates genuine network TIs. We move beyond regular lattices to complex networks spanning regular, small-world, and random architectures. To separate the node-degree effect identified in Fig. 2 from network connectivity, we construct degree-$k$ networks and tune their architecture using the degree-preserving swap parameter $\rho$ (Supplementary

Note S6): $\rho = 0$ for regular networks, $\rho \approx 0.1$ for small-world networks, and $\rho \approx 10$ for random networks (Fig. 3a). All networks are initialized with random hopping phases.

To transplant 2D TI physics into these abstract networks, we randomly project each network onto a plane (Fig. 3b,c) and designate the outermost 40 % of nodes as edge nodes, guided by the corresponding ratio in the standard Hofstadter model. This configuration maps the protected directional transport envisioned in Fig. 1a onto topological edge-state transport. Under our proposal—topological heterogeneity—we apply the GDS-JDS policy to jointly optimize $\{\tau_p\} = \{\{\varphi_{mn}\}, \{(x_n, y_n)\}\}$, targeting the bulk $C_L^0 = 1$ (Fig. 3d) and a large bulk $\mu_L$.

The resulting *k*-*ρ* maps reveal topological phases across the landscape of complex networks. We evaluate the ensemble averaged bulk quantities $\langle C_L \rangle$ (Fig. 3e) and $\langle \mu_L \rangle$ (Fig. 3f), together with the empirical edge-confinement measure[42] $\langle \gamma \rangle$ (Fig. 3g; see Methods). At sufficiently high *k*, the optimized networks achieve a nontrivial bulk with $\langle C_L \rangle = 1$ across the regular, small-world, and random regimes (Fig. 3e). Within this topological regime, edge confinement closely tracks the opening of a sizable local gap (Fig. 3g versus 3f). This observation highlights the critical role of a large bulk $\mu_L$ in complex networks with nonlocal hopping, for which conventional bulk-edge correspondence does not explicitly guarantee the existence of edge states[21].

We perform extended analyses of Hofstadter butterflies (Supplementary Note S6), optimization curves (Supplementary Note S7), localization and graph-theoretic characteristics (Supplementary Note S8), spectral responses (Supplementary Notes S9 and S10), and flux distributions (Supplementary Note S11). These analyses support the emergence of nontrivial topology in genuine complex networks and further confirm the Haldane-type nature at high node degree. However, these diagnostics do not by themselves establish topologically protected transport. We therefore next verify the chirality and disorder robustness of the edge-confined states.

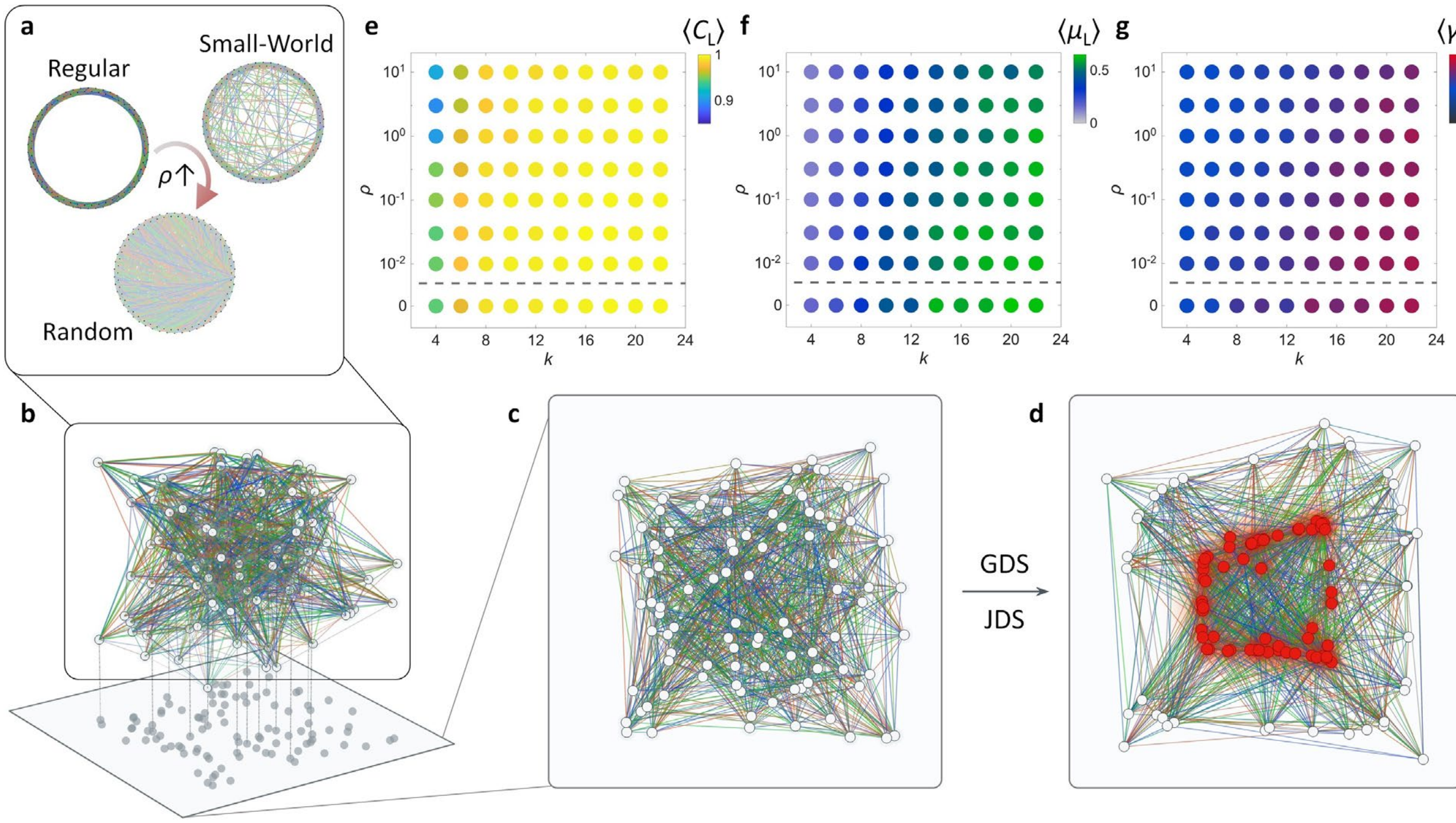

**Fig. 3. Topological phases across complex networks**. **a,** Degree-$k$ networks evolve from regular to random structures with increasing $\rho$. **b,** Projection of a network onto a 2D plane. **c,d,** Projected network before (**c**) and after (**d**) optimization; red nodes denote $C_L = 1$. **e-g,** $k$-$\rho$ maps of $\langle C_L \rangle$ (**e**), $\langle \mu_L \rangle$ (**f**), and $\langle \gamma \rangle$ (**g**). Dashed lines separate the regular case ($\rho = 0$) from rewired networks.

### Small-worldness in network TIs

Although every node in our networks has the same degree, rewiring creates a pronounced hierarchy in node importance (Supplementary Note S6). We quantify this hierarchy using betweenness centrality[43]—the fraction of shortest paths between node pairs that traverse a given node—thereby distinguishing hubs from peripheral nodes (Fig. 4a). This emergent centrality landscape provides a network-specific axis for determining how network topology reshapes topological wave transport.

To test this idea, we develop a network router based on an overcoupled five-port geometry[44] (Fig. 4b,c; Supplementary Note S12 for weakly coupled excitation). The optimized network TIs exhibit pronounced chiral circulation, quantified by the chirality parameter $\eta$ (Methods), which peaks near the target energy $E = 1$ (Fig. 4d; Supplementary Note S13). The designed transport therefore inherits the handedness of chiral edge states, realizing directional transport through complex networks. Around this energy, regular and small-world networks retain markedly higher chirality than random networks—underscoring the critical role of network clustering in chiral transport.

To investigate another critical feature—disorder immunity—we focus on small-world TIs, which combine regular-like chirality with strongly differentiated node centralities. We perturb the onsite energies (Fig. 4e), hopping amplitudes (Fig. 4f), and hopping phases (Fig. 4g) associated with either hubs or peripheral nodes of the $\eta$-optimal design (Methods). A striking inversion of sensitivity emerges: onsite disorder is more detrimental at peripheral nodes, whereas hopping-amplitude and, most prominently, hopping-phase disorder suppress chirality more strongly when applied to hubs. Topological robustness is therefore not uniformly distributed across a complex network but jointly encoded by where disorder occurs and how it enters the Hamiltonian. Betweenness hubs can tolerate onsite disorder yet become Achilles' heels under link disorder, revealing a node- and perturbation-selective vulnerability of topological transport without a counterpart in prior studies.

For comparison, we evaluate the robustness against disorder of regular and random TIs in Supplementary Note S14. These results demonstrate superior chiral robustness in small-world and random networks, both characterized by a short characteristic path length. Together with Fig. 4d, this observation highlights the decisive role of network topology in governing topological wave transport, emphasizing the promise of small-world TIs for achieving high chirality, enhanced robustness, and reconfigurability.

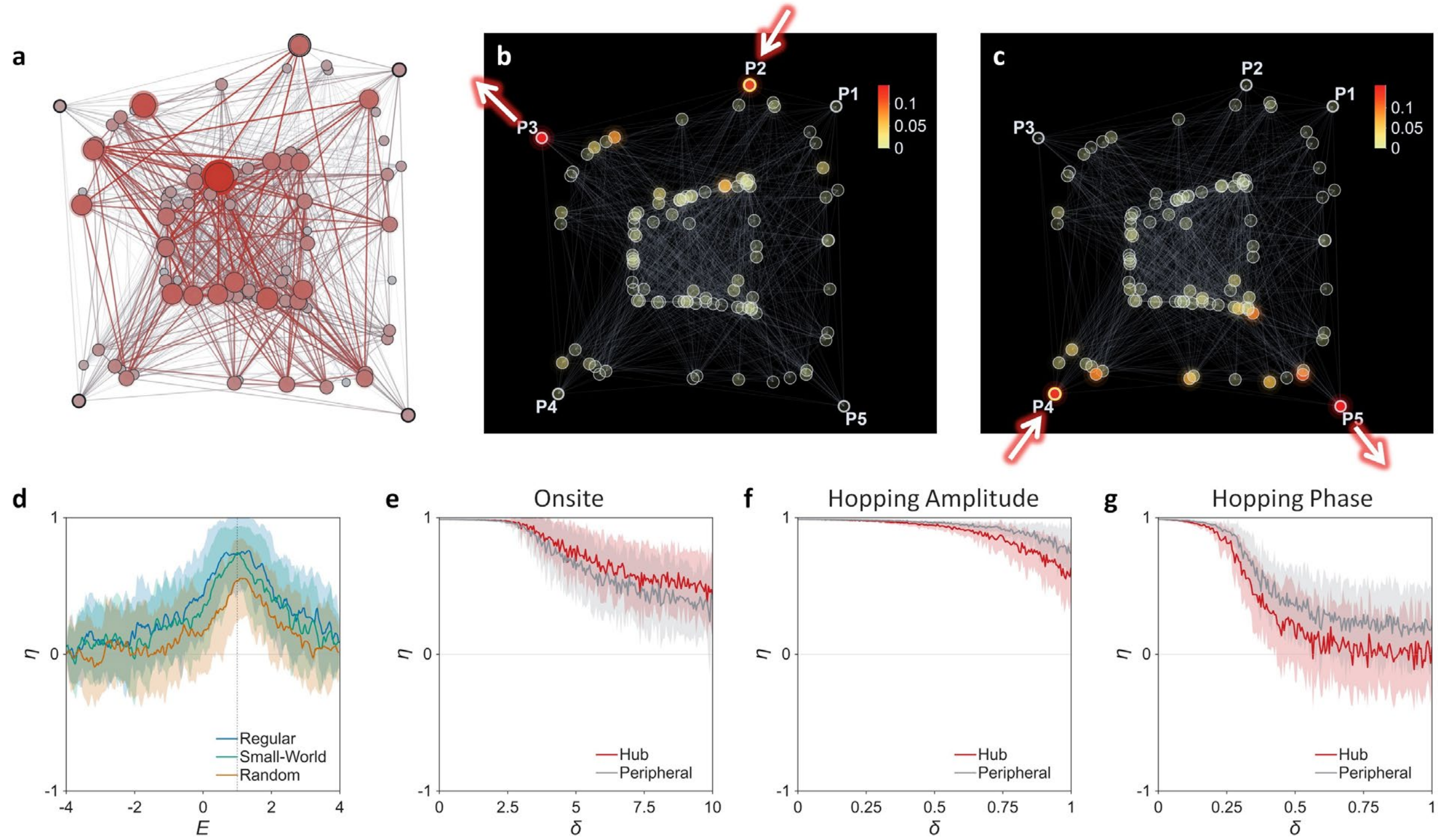

**Fig. 4. Small world provides selective topological robustness**. **a,** Betweenness-centrality landscape of a degree-$k$ small-world TI. Node size and colour represent centrality. **b,c,** Chiral transport in the scattering geometry; arrows mark incident and outgoing channels, and node colours show intensity. The ports are strongly coupled to the network with $t_{\text{ext}} = 2.49 > t_{mn} = 1$ (Supplementary Movies S1 and S2 for **b** and **c**, respectively). **d,** Spectrally-resolved chirality $\eta(E)$. The vertical dashed line marks $E = 1$. **e-g,** $\eta(E = 1)$ under onsite (**e**), hopping-amplitude (**f**), and hopping-phase (**g**) perturbations of strength $\delta$ applied to the most central half (hubs, red) or the least central half (peripheral nodes, grey) of the network, at $t_{\text{ext}} = 2.67$. Curves and shaded regions denote ensemble averages and one standard deviation, respectively, for 40 disorder realizations.

## Discussion

We validated our core design principle—realizing a robust, topologically nontrivial bulk to support protected transport—through multiple complementary approaches: spectral localizers, spectrally resolved analyses of edge-state confinement and localization, and analysis of scattering spectra over various network topologies. These rigorous validations motivate extending our approach beyond 2D Hermitian topology. Specifically, our multi-objective design using the Hellmann-Feynman theorem[38], guided by surrogate objectives for topological invariants, could be directly adapted to one-dimensional, higher-dimensional, and non-Hermitian systems using spectral localizers appropriate to their symmetries and gap structures[21]. Such extensions could facilitate the exploration of higher-order topological insulators[45], Yang monopoles[46], and non-Hermitian skin effects[47] in complex networks.

Our design uses 2D embedding of abstract networks ranging from regular to random. The resulting chiral transport and its selective control can therefore be reproduced in physical architectures of different spatial dimensions and embedding, provided that the designed Hamiltonian and input-output couplings are preserved. This platform transparency suggests opportunities for realizing robust transport with long-range hopping in synthetic-dimensional networks[48], network lasers[49], quantum electrodynamics[30], and scattering systems[50]. Our findings also raise an intriguing question: whether existing biological, computational, or macroscopic networks can admit magnetic-graph descriptions and may already support analogous topological phases.

In conclusion, we have demonstrated TI phases in complex networks. Network characteristics—node degree, randomness, and shortcuts—emerge as key parameters governing the topological properties of waves. While practical implementation requires the realization of large-scale magnetic graphs with controlled hopping amplitudes and phases, our results motivate the exploration of related phenomena in quantum networks[30], higher-dimensional photonic networks[51], and artificial neural networks[52].

## Methods

**Spectral localizer analysis.** To diagnose topological properties of finite 2D systems, we employ the spectral localizer[21,22]:

$$L(n,E;\kappa)=\left(H-EI\right)\otimes\sigma_z+\kappa\left[\left(X-x_nI\right)\otimes\sigma_x+\left(Y-y_nI\right)\otimes\sigma_y\right], \tag{3}$$

where $I$ is the identity matrix, $X$ and $Y$ are the position operators, and $\sigma_{x,y,z}$ are the Pauli matrices. Acting on a doubled Hilbert space, the spectral localizer combines the energy shift $H - EI$ with the spatial shifts $X - x_nI$ and $Y - y_nI$ into a single Hermitian operator $L$. The local topology can be characterized by three quantities: $C_L$ and $\mu_L$ in Eq. (2), and the quadratic local gap $\mu_Q$, defined as

$$\mu_Q(n,E;\kappa)=\sqrt{\min\left[\left|\mathrm{Spec}\left(Q\right)\right|\right]}, \tag{4}$$

where $Q(n, E; \kappa) = \kappa^2[(X - x_nI)^2 + (Y - y_nI)^2] + (H - EI)^2$ is a quadratic composite operator that quantifies the joint spatial and spectral proximity of a state to the $n$-th site and the target energy $E$. At a position-energy point where $C_L$ changes, the gap closes, $\mu_L \to 0$, by definition[21]. When the Hamiltonian is sufficiently local, a real-space manifestation of bulk-edge correspondence is expressed by the inequality[21]:

$$\left|\left(\mu_L(n,E;\kappa)\right)^2-\left(\mu_Q(n,E;\kappa)\right)^2\right|\le\sum_{j=1}^{d}\kappa\left\|\left[X,H\right]+\left[Y,H\right]\right\|, \tag{5}$$

which yields small $\mu_Q$—indicating the existence of a state—using the small value of $\|[X, H] + [Y, H]\|$ and $\mu_L \to 0$. In complex networks with strongly nonlocal connectivity, however, these commutator norms can become large, making the bound too loose to provide a useful guarantee[27]. Therefore, we tailor $C_L$ and $\mu_L$ in exploring topological insulators in complex networks, focusing on achieving a robust nontrivial bulk with $C_L \neq 0$ and large $\mu_L$.

**Gradients for the inverse design.** We apply the Hellmann-Feynman theorem[38] to the target differentiable metrics $C_\beta$ and $\mu_L$, achieving

$$\nabla C_\beta = \frac{\beta}{2}\sum_{j=1}^{2N}\text{sech}^2\left(\beta\lambda_j\right)\left\langle u_j\left|\nabla L\right|u_j\right\rangle, \quad \nabla\mu_{\text{L}} = \text{sgn}\left(\lambda_g\right)\left\langle u_g\left|\nabla L\right|u_g\right\rangle, \tag{6}$$

where $|u_j\rangle$ and $\lambda_j$ denote the $j$-th eigenstate and eigenvalue of $L$, respectively, with the eigenvalues ordered increasingly, and $g$ labels the eigenvalue of $L$ with the smallest absolute value. The spectral-localizer gradient $\nabla L$ required in Eq. (6) can be obtained from Eq. (3) using Eq. (1).

**Edge confinement.** To quantify how strongly the eigenstate at the design energy is confined to the edge, we employ the edge-weight fraction used as a heuristic measure in disordered[42] and hyperbolic[5] topological systems:

$$\gamma = \sum_{r\in\Lambda_{\text{edge}}}\left|\psi_r\right|^2, \tag{7}$$

where $\psi_r$ is the amplitude at node $r$ of the normalized eigenstate whose eigenenergy is closest to $E = 1$, and $\Lambda_{\text{edge}}$ is the set of edge nodes.

**Chirality parameter.** With $N_p$ ports attached to the boundary of the network and indexed counterclockwise, the port-to-port transmission is $T_{pq}(E) = |S_{pq}(E)|^2$, where $S_{pq}$ is the scattering amplitude from input port $q$ to output port $p$. One-way circulation is quantified by the counterclockwise and clockwise adjacent-port transmissions, as follows:

$$T_{\text{CCW}}\left(E\right) = \frac{1}{N_p}\sum_{q=1}^{N_p}T_{q+1,q}\left(E\right), \qquad T_{\text{CW}}\left(E\right) = \frac{1}{N_p}\sum_{q=1}^{N_p}T_{q-1,q}\left(E\right). \tag{8}$$

The chirality parameter is defined as

$$\eta\left(E\right) = \frac{T_{\text{CCW}}\left(E\right) - T_{\text{CW}}\left(E\right)}{T_{\text{CCW}}\left(E\right) + T_{\text{CW}}\left(E\right)}, \tag{9}$$

which ranges from $-1$ to $+1$: $\eta = +1$ (or $-1$) for a perfect counterclockwise (or clockwise) circulator.

**Perturbations**. For the device of Fig. 4, nodes are ranked by betweenness centrality and links by the rank of their less central end. The hub (or peripheral) group is the most (or least) central half of the nodes for onsite perturbations, or of the links for hopping perturbations. Independent values drawn uniformly from $[-\delta, \delta]$ are added to the onsite energies, hopping amplitudes, or hopping phases of the selected group, with all other parameters, node positions, and ports unchanged. $\eta(E = 1)$ is evaluated from the five-port scattering matrix and averaged over 40 perturbation realizations per $\delta$.

## Data availability

Data used in the current study are available from the corresponding authors upon request and can also be obtained by running the shared codes at 10.5281/zenodo.23095572 in the Zenodo[53].

## Code availability

All original code has been deposited at Zenodo[53] and is publicly available at 10.5281/zenodo.23095572.

## Acknowledgements

We acknowledge financial support from the National Research Foundation of Korea (NRF) through the Core Research Grants (no. RS-2026-25469085), Innovation Research Center (no. RS-2024-00413957), Young Researcher Program (no. RS-2025-00552989), Pilot and Feasibility Grants (no. RS-2025-19912971), and Midcareer Researcher Program (no. RS-2023-00274348), all funded by the Korean government. This work was supported by the BK21 FOUR program of the Education and Research Program for Future ICT Pioneers in 2026, through Seoul National University. We also acknowledge an administrative support from SOFT foundry institute.

## Author Contributions

S.Y., X.P., and N.P. contributed equally to conceiving the idea, developing the code, discussing the results, and preparing the final manuscript.

## Competing Interests

The authors have no conflicts of interest to declare.

## Additional information

**Correspondence and requests for materials** should be addressed to S.Y., X.P., or N.P.

## Figure Legends

**Fig. 1. Designing network TIs**. **a,** Small-world TI supporting a protected directional state (blue glows and arrows)—the network analogue of chiral edge states in 2D TIs. The realization is one instance of the designs in Fig. 3, for $k = 20$ and link-swap parameter $\rho = 0.1$. **b,** Surrogate model $C_\beta$ for $C_\mathrm{L}$ across the design space $\tau$. **c,** Optimization curves for **a**.

**Fig. 2. High-degree Hofstadter lattices enhance information capacity**. **a,** Hopping shells. **b,** Optimization of a 4NN network toward $C_\mathrm{L}^0 = 3$. Node and link colours indicate $C_\mathrm{L}$ and the directed hopping phase (grey for 0, orange for $\pi/2$, green for $\pi$, and blue for $3\pi/2$). **c,** Network Hofstadter butterfly. Points denote eigenenergies. $\beta = 1$ in **b** and **c**. **d,e,** Ensemble-averaged bulk $C_\mathrm{L}$ (**d**) and $\mu_\mathrm{L}$ for bulk (filled) and edge (open) nodes (**e**): 50 realizations per condition. Blue circles, orange triangles, and green diamonds denote $C_\mathrm{L}^0 = 1$, 2, and 3, respectively. Faint points show individual realizations; error bars, one standard deviation.

**Fig. 3. Topological phases across complex networks**. **a,** Degree-$k$ networks evolve from regular to random structures with increasing $\rho$. **b,** Projection of a network onto a 2D plane. **c,d,** Projected network before (**c**) and after (**d**) optimization; red nodes denote $C_\mathrm{L} = 1$. **e-g,** $k$-$\rho$ maps of $\langle C_\mathrm{L} \rangle$ (**e**), $\langle \mu_\mathrm{L} \rangle$ (**f**), and $\langle \gamma \rangle$ (**g**). Dashed lines separate the regular case ($\rho = 0$) from rewired networks.

**Fig. 4. Small world provides selective topological robustness**. **a,** Betweenness-centrality landscape of a degree-$k$ small-world TI. Node size and colour represent centrality. **b,c,** Chiral transport in the scattering geometry; arrows mark incident and outgoing channels, and node colours show intensity. The ports are strongly coupled to the network with $t_\mathrm{ext} = 2.49 > t_{mn} = 1$ (Supplementary Movies S1 and S2 for **b** and **c**, respectively). **d,** Spectrally-resolved chirality $\eta(E)$. The vertical dashed line marks $E = 1$. **e-g,** $\eta(E = 1)$ under onsite (**e**), hopping-amplitude (**f**), and hopping-phase (**g**) perturbations of strength $\delta$ applied to the most central half (hubs, red) or the least central half (peripheral nodes, grey) of the network, at $t_\mathrm{ext} = 2.67$. Curves and shaded regions denote ensemble averages and one standard deviation, respectively, for 40 disorder realizations.

# Supplementary Information for "Topological insulators on complex networks"

Sunkyu Yu[1†], Xianji Piao[2§], and Namkyoo Park[3*]

[1]Intelligent Wave Systems Laboratory, Department of Electrical and Computer Engineering, Seoul National University, Seoul 08826, Korea

[2]Wave Engineering Laboratory, School of Electrical and Computer Engineering, University of Seoul, Seoul 02504, Korea

[3]Photonic Systems Laboratory, Department of Electrical and Computer Engineering, Seoul National University, Seoul 08826, Korea

E-mail address for correspondence: [†]sunkyu.yu@snu.ac.kr, [§]piao@uos.ac.kr, [*]nkpark@snu.ac.kr

## Supplementary Note S1. Jacobian-descent scheme

In this note, we describe the multi-objective update rule, which is used in the second stage of every design in the main text. Our optimization targets the simultaneous minimization of two cost functions: the Chern-index cost $K_{\mathrm{C}}$ and the local-gap cost $K_{\mu}$. $K_{\mathrm{C}}$ penalizes the deviation of the surrogate $C_{\beta}$ from its target $C_{\mathrm{L}}^{0}$ on the cost nodes: the bulk nodes for the design in Fig. 2 and all nodes for the designs in Figs. 3 and 4 of the main text. On the other hand, $K_{\mu}$ penalizes violations of the bounds imposed on $\mu_{\mathrm{L}}$: a lower bound on the bulk nodes and, for the designs in Figs. 3 and 4, an upper bound on the edge nodes. Because the two costs depend on the same design variables, their gradients $\nabla K_{C}$ and $\nabla K_{\mu}$ are generally not aligned, as exemplified by the channel-bandwidth tradeoff in topological systems [1,2].

To guarantee the simultaneous descent of all costs, we employ a dual-cone projection [3,4] of the gradients (Fig. S1). Let $g_d = \nabla K_d$ ($d = C, \mu$) denote the gradients of the cost functions with respect to the design-parameter vector $\tau = [\tau_1, \tau_2, \ldots, \tau_{\mathrm{M}}]^{\mathrm{T}}$ and let $G = [g_C\, g_\mu]$ be the matrix collecting them. The dual cone

$$\Omega = \left\{ \upsilon : G^{\mathrm{T}} \upsilon \geq 0 \right\} = \left\{ \upsilon : g_d \cdot \upsilon \geq 0 \text{ for all } d \right\}, \tag{S1}$$

is the set of directions along which every cost is nondecreasing to first order. Therefore, for any $\tau \in \Omega$, $-\tau$ is a common descent direction for all costs. To exploit the property of the dual cone, we project each gradient onto $\Omega$ through its Moreau decomposition,

$$\Pi_{\Omega}\left(g_d\right) = g_d + G a_d, \qquad a_d = \underset{a \geq 0}{\operatorname{argmin}} \left\| G a + g_d \right\|, \tag{S2}$$

which is a small nonnegative least-squares problem that we solve exactly by enumerating the active sets of the constraint $a \geq 0$. The design variables are then updated along with the weighted sum of the projected gradients,

$$\Delta\tau = -s\sum_{d}\omega_{d}\Pi_{\Omega}\left(g_{d}\right), \tag{S3}$$

where $\omega_d$ are task weights and $s$ is the step size. Because every $\Pi_\Omega(g_d)$ lies in $\Omega$, we have $g_d \cdot \Delta\tau \leq 0$ for all $d$, achieving the nonincreasing condition of all costs to first order.

In all designs, the optimization proceeds in two stages. In the first stage, a GDS minimizes only $K_C$ by steepest descent until $K_C$ falls below $10^{-3}$ or 300 updates have been accepted. In the second stage, the JDS of Eq. (S3) minimizes $K_C$ and $K_\mu$ jointly with unit task weights ($\omega_C = \omega_\mu = 1$) and cost normalization factors of 1 and 0.5 for $K_C$ and $K_\mu$, respectively, under the same stopping rule as in the first stage, requiring both costs to fall below $10^{-3}$. The step size $s$ is determined at every update by an Armijo backtracking line search, and a trial step is accepted only when it satisfies the Armijo sufficient-decrease condition for every cost scheduled in the stage.

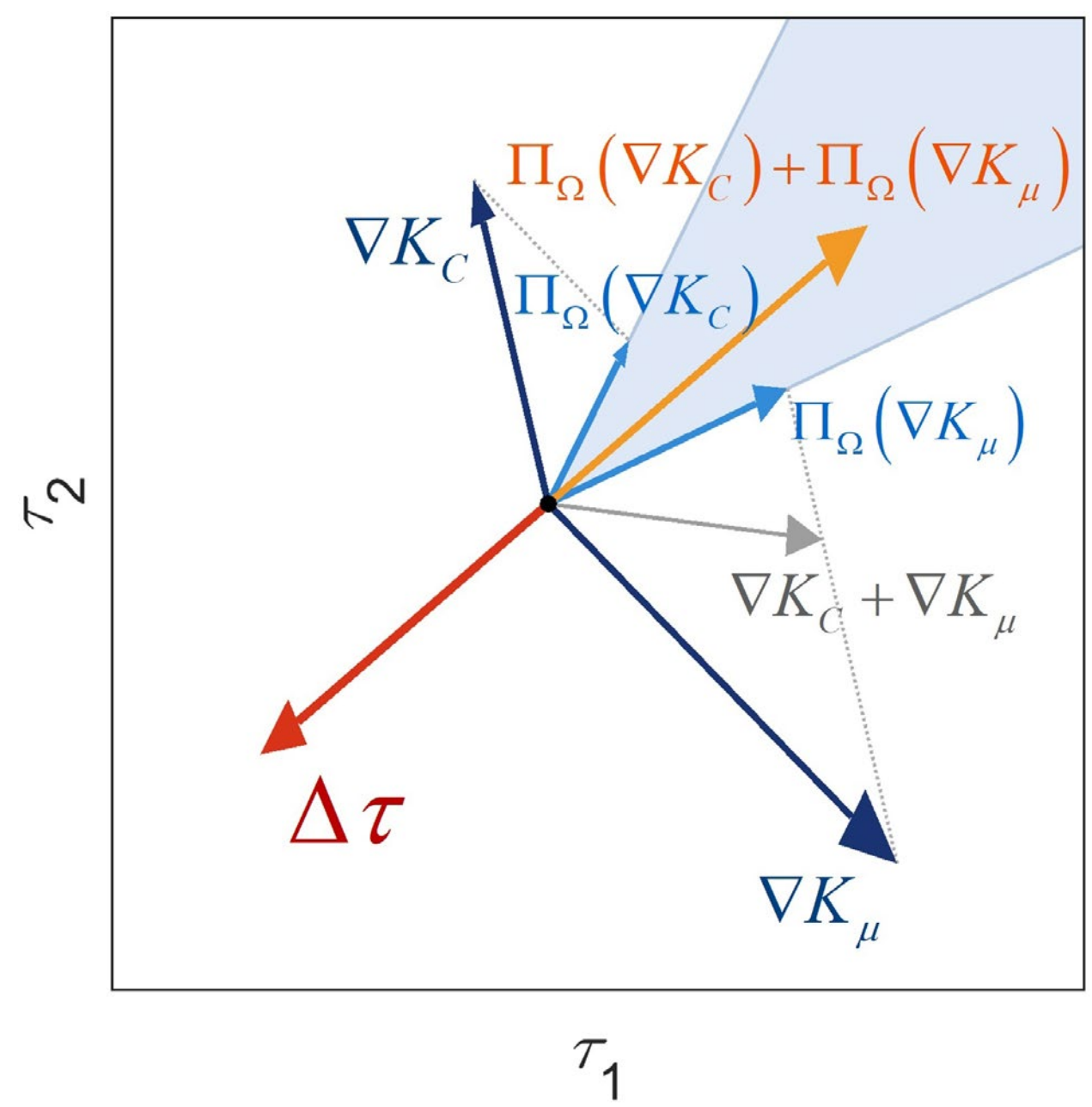

**Fig. S1. Jacobian-descent update for conflicting objectives.** Two conflicting cost gradients $\nabla K_C$ and $\nabla K_\mu$ (dark blue arrows) in a 2D design space. $\Pi_\Omega(\nabla K_C)$ and $\Pi_\Omega(\nabla K_\mu)$ (light blue arrows) denote their projections onto the dual cone $\Omega$ (shaded region). Orange and red arrows denote the weighted sum $\Pi_\Omega(\nabla K_C) + \Pi_\Omega(\nabla K_\mu)$ and the resulting update $\Delta\tau$, respectively.

**Supplementary Note S2. Standard Hofstadter model**

The lattice designs in Fig. 2 of the main text are benchmarked against the standard Hofstadter model [5], whose local topological characteristics are shown in Figs. S2 and S3. We consider an $11 \times 11$ square lattice with $N = 121$ nodes, a uniform NN hopping amplitude $t_{mn} = 1$, and zero on-site energy $\varepsilon_n = 0$. The NN hopping phases $\varphi_{mn}$ are assigned in the Landau gauge, yielding a uniform flux $\theta$ across the system, as determined by the Peierls phase per plaquette.

Figure S2 shows an example of the real-space distributions of the local Chern index $C_L$ and the local gap $\mu_L$ at $\theta = 3\pi/2$. Notably, the $C_L$ distribution separates the bulk—the 81 interior nodes with $C_L = +1$—from the edge—the 40 boundary nodes with $C_L = 0$ (Fig. S2a). Correspondingly, $\mu_L$ is finite and roughly uniform in the bulk and small on the edge except at the four corner nodes (Fig. S2b), as expected because the transition of $C_L$ across the boundary requires the local gap closing (Fig. S2b). We employ this bulk-edge separation in all lattice designs, targeting a nontrivial $C_L = C_L{}^0$ and a large $\mu_L$ on the bulk nodes.

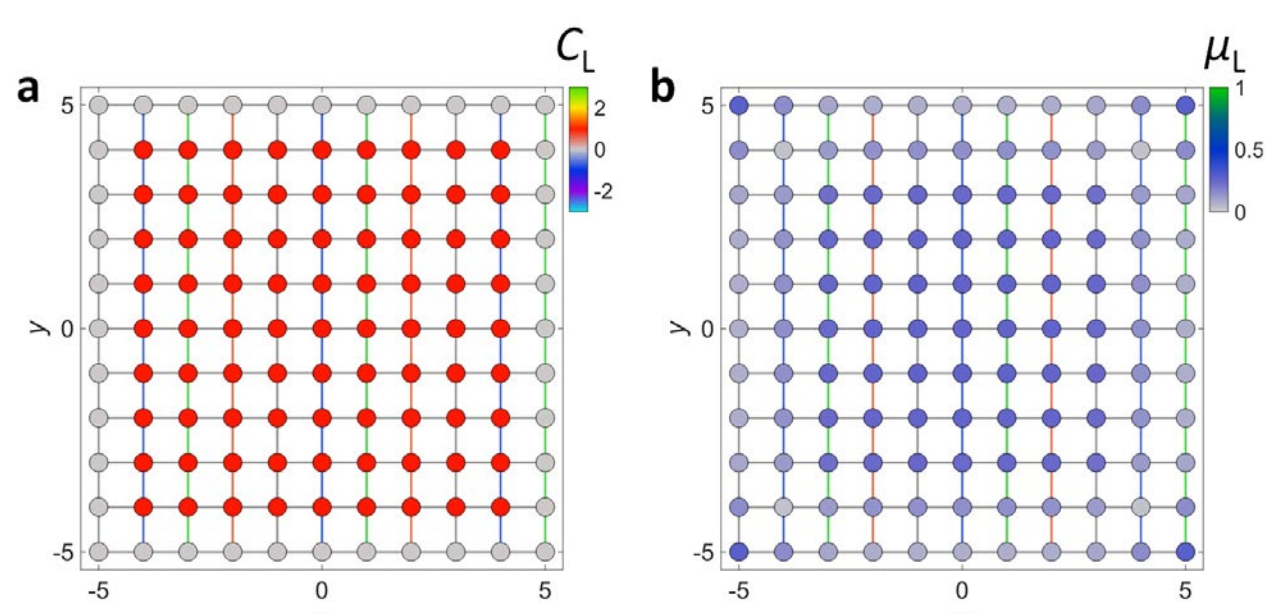

**Fig. S2. Hofstadter reference lattice. a**, $C_L$ and **b**, $\mu_L$ evaluated at $E = 1$ and $\kappa = 0.25$. Link colours encode the directed hopping phase (grey for 0, orange for π/2, green for π, and blue for 3π/2).

Figure S3 shows the Hofstadter butterflies of the same lattice, plotted for the bulk- and edge-averaged $C_L$ and $\mu_L$ as functions of the plaquette phase $\theta$ and the probe energy $E \in [-4, 4]$, with the eigenenergies overlaid. The emergence of topologically nontrivial gaps and their bulk-edge contrast are confirmed by the bulk averages (Figs. S3a and S3b) and the corresponding edge averages (Figs. S3c and S3d). At the energy level of interest in our work, $E = 1$ (black dashed lines), the accessible local Chern index is limited to $|C_L| \leq 2$, and the $|C_L| = 2$ regions are highly fragile as shown in their much smaller local gaps (Fig. S3b). Therefore, the only value that is practically meaningful in terms of topological protection in the standard Hofstadter model is $|C_L| = 1$, which eventually limits the number of accessible topologically protected edge channels.

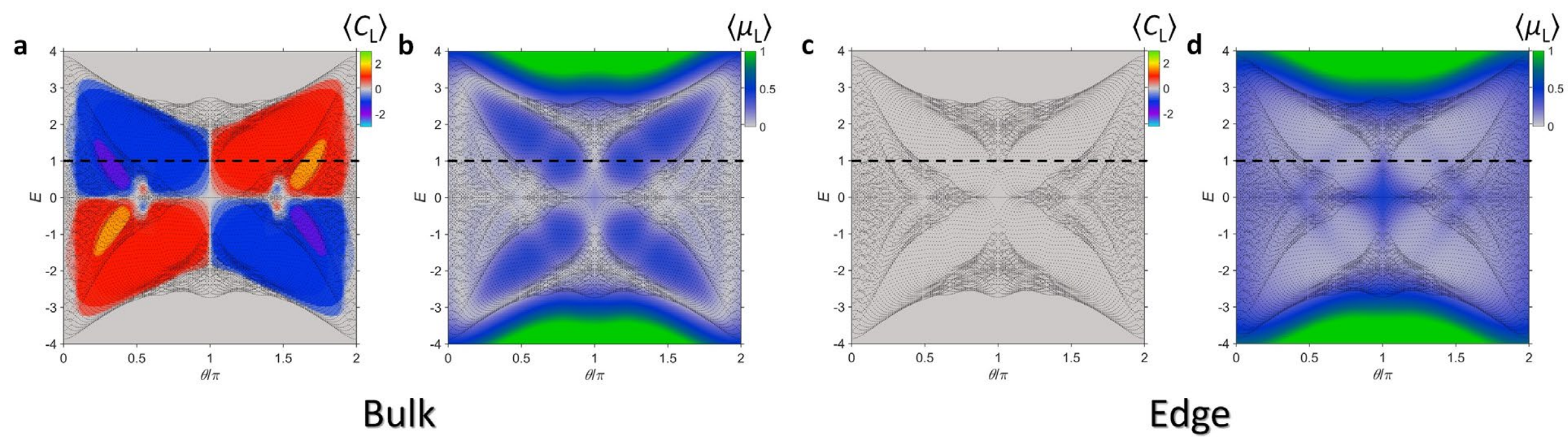

**Fig. S3. Butterflies of the Hofstadter reference. a,b**, Bulk-averaged $C_L$ (**a**) and $\mu_L$ (**b**) of the lattice in Fig. S2 as functions of the plaquette phase $\theta$ and energy $E$. **c,d**, The same quantities averaged over the edge nodes. Black points denote the eigenenergies; dashed lines mark the design energy $E = 1$.

**Supplementary Note S3. Extended data for *k*-degree Hofstadter networks**

Figure S4 complements Fig. 2b in the main text by showing the corresponding local gap $\mu_L$ before and after the optimization. For the random initial phases, $\mu_L$ is small especially near the centre of the network. After the optimization, $\mu_L$ achieves a bulk-to-edge gradient: it reaches $\mu_L \approx 1$ at the centre and gradually drops to $\mu_L \approx 0.1$ on the edge except for four corners. The coexistence of a large interior gap with a small edge gap is consistent with the Hofstadter reference (Fig. S2b), while the maximum value of $\mu_L$ substantially increases in the 4NN case.

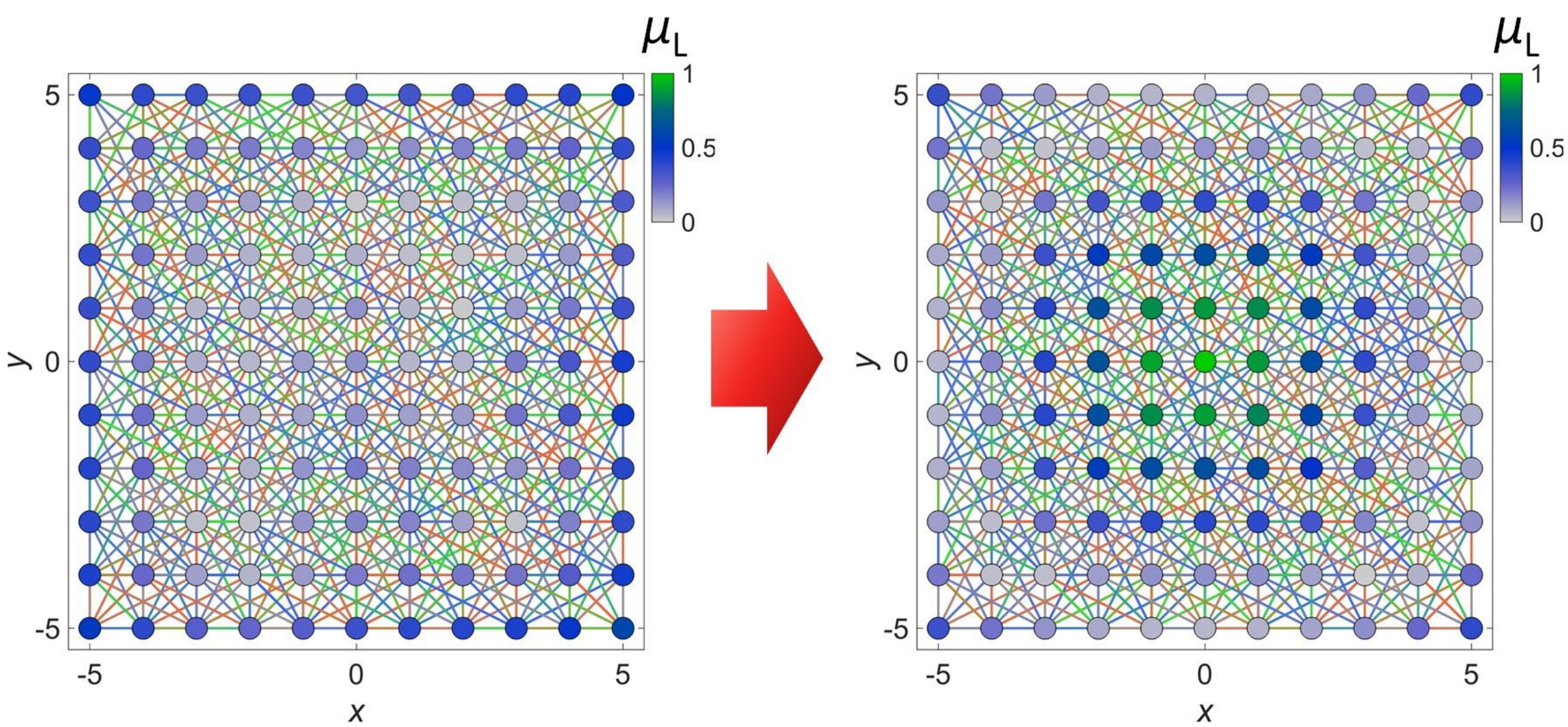

**Fig. S4. Local gap before and after optimization.** $\mu_L$ maps of the 4NN lattice of Fig. 2b ($C_L^0 = 3, \beta = 1$) before (left) and after (right) optimization of the hopping phases. Link colours encode the hopping phases as in Fig. S2. Node colours indicate $\mu_L$. All other parameters are the same as those in Fig. 2b.

Because our framework optimizes the system parameters $\{\tau_p\}$ directly for the target $C_L$ and $\mu_L$, the resulting design does not guarantee a uniform plaquette flux $\theta$, even in the NN configuration. Furthermore, the NNN, 3NN, and 4NN configurations preclude a straightforward definition of the plaquette flux because of substantial link crossings. Therefore, to examine how the spectral and topological characteristics respond to a change of the magnetic features, we introduce an alternative approach: rescaling every hopping phase by a common scalar factor $\chi$ while keeping the other system parameters fixed:

$$
\begin{aligned}
\left[H(\chi)\right]_{pq} &= \left|\left[H\right]_{pq}\right| \exp\left(i\chi\angle\left[H\right]_{pq}\right) \qquad (p<q) \\
\left[H(\chi)\right]_{pq} &= \left|\left[H\right]_{pq}\right| \exp\left(-i\chi\angle\left[H\right]_{qp}\right) \qquad (p>q),
\end{aligned}
\tag{S4}
$$

where ∠ denotes the principal argument in (−π, π]. Because the flux through a cycle is the sum of the directed link phases around it, Eq. (S4) multiplies the flux through every cycle of the network by $\chi$: $\chi = 1$ for the optimized design, $\chi = 0$ for the flux-free network, and $\chi = 2$ for the system with doubled fluxes.

Figure S5 shows the corresponding network Hofstadter butterflies of one realization of each of the four designs of Fig. 2 in the main text (NN, NNN, 3NN and 4NN with $C_L^0 = 1, 1, 2,$ and 3, respectively). Every design exhibits a plateau of the bulk-averaged $C_L$ at or close to its target value around the design point: $\chi = 1$ and $E = 1$ (Figs. S5a,e,i,m). These plateaus coincide with distributions of enlarged bulk-averaged local gaps (Figs. S5b,f,j,n), confirming the protection of the designed $C_L$. The edge-averaged quantities ($C_L$ in Figs. S5c,g,k,o and $\mu_L$ in Figs. S5d,h,l,p) also confirm the transition of the local topology from the bulk to the edge.

We note that the NN design shares the characteristics of the standard Hofstadter butterflies in Fig. S3, such as the mirror symmetry about $E = 0$, which originates from the chiral symmetry of the bipartite NN lattice. In contrast, the longer-range hopping amplitudes of the NNN, 3NN, and 4NN designs connect sites of the same sublattice and thus break this symmetry, leaving a single $C_L$ lobe along the $E$ axis (Figs. S5e,i,m).

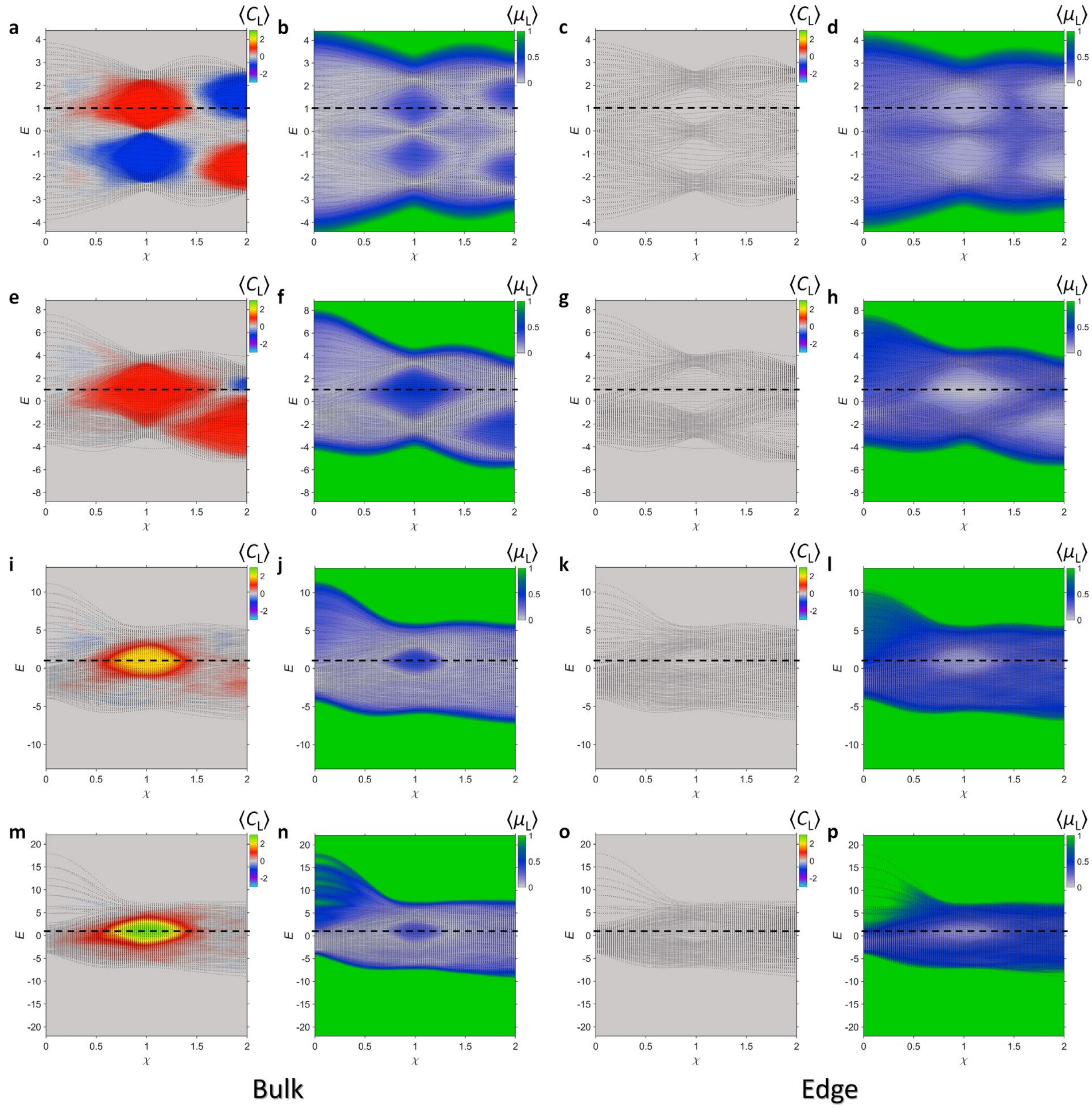

**Fig. S5. Network Hofstadter butterflies of the optimized lattices. a-d**, NN, **e-h**, NNN, **i-l**, 3NN, and **m-p**, 4NN designs with $C_L{}^0 = 1$, 1, 2, and 3, respectively ($\beta = 1$; the 4NN case is that of Figs. 2b,c: bulk-averaged $C_L$ (**a,e,i,m**) and $\mu_L$ (**b,f,j,n**), and edge-averaged $C_L$ (**c,g,k,o**) and $\mu_L$ (**d,h,l,p**), as functions of the phase-rescaling factor $\chi$ and energy $E$. Points denote eigenenergies; dashed lines mark $E = 1$. All other parameters are the same as those in Fig. 2c.

**Supplementary Note S4. Design via a higher-fidelity surrogate**

In our design, the fidelity of the $C_L$ surrogate, $C_\beta$, plays a critical role in exploring the design space. A smaller $\beta$, indicating a smoother $C_\beta$, guarantees more stable gradients, at the expense of the fidelity between $C_\beta$ and $C_L$. Therefore, $\beta$ controls the tradeoff between the fidelity of the surrogate and the smoothness of the cost landscape.

Figure S6 repeats the analysis of Fig. 2d,e in the main text for $\beta = 10$ instead of $\beta = 1$. The target $C_L^0 = 1$ is reached for all hopping ranges at $\beta = 10$, and for the NN and NNN lattices at $\beta = 1$. The two values of $\beta$ differ in how they handle the higher $C_L^0$ targets. The smoother surrogate of $\beta = 1$ reaches higher $C_L$ ($C_L^0 = 3$; Fig. 2d in the main text) because its gradients remain finite over a wide range of the localizer spectrum. However, it often loses control over the target value: for the 3NN and 4NN lattices, $C_L^0 = 1$ is overshot to $C_L \approx 2$, and for the 4NN lattice, $C_L^0 = 2$ is overshot to $C_L \approx 3$. On the other hand, the sharper surrogate ($\beta = 10$) instead respects the target— $C_L$ does not exceed $C_L^0$ on any bulk node in any of the 600 realizations (Fig. S6a)—whereas $C_L^0 = 3$ is no longer reached.

Although $\beta$ therefore operates as a hyperparameter that trades reachability against controllability, both choices yield a wider design space than the standard Hofstadter model in terms of $C_L$ (Fig. S6a and Fig. 2d in the main text) and $\mu_L$ (Fig. S6b and Fig. 2e in the main text).

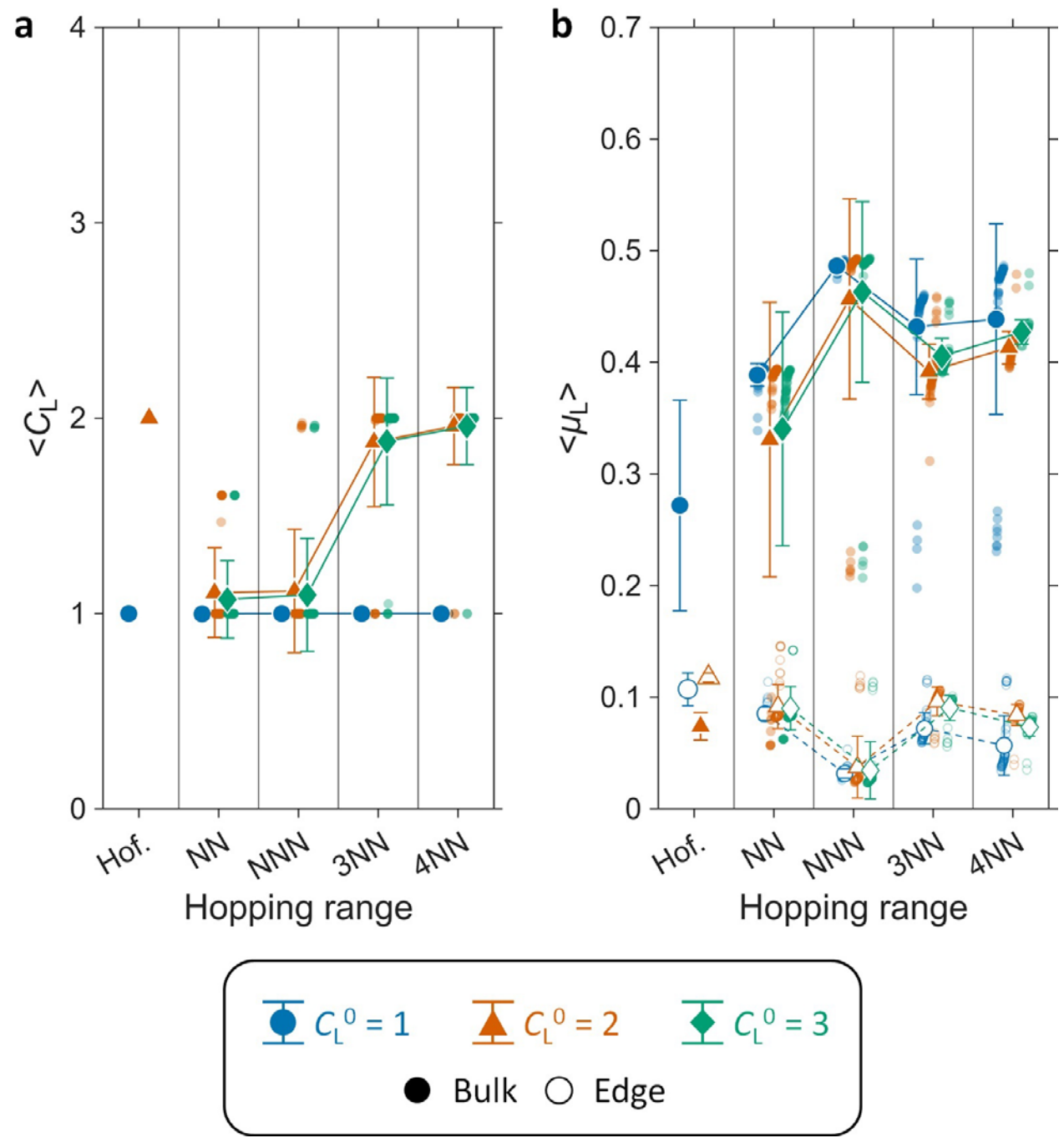

**Fig. S6. Ensemble statistics at *β* = 10. a,b**, Ensemble-averaged bulk $C_L$ (**a**) and $\mu_L$ for bulk (filled) and edge (open) nodes (**b**) versus hopping range, for optimizations with $\beta = 10$ (50 realizations per condition). All other parameters are the same as those in Figs. 2d,e.

**Supplementary Note S5. Flux evaluations in high-degree Hofstadter networks**

In contrast to the standard Hofstadter lattice, it is not straightforward to define a plaquette flux in high-degree Hofstadter networks and in the network topological insulators due to link crossings: the plaquettes of the standard model are the faces of a planar embedding, which form a cycle basis of a planar graph [6], whereas the geometric embedding of a graph with crossing links has no such faces. As a generalization that contains the standard Hofstadter model as a special case, we measure the flux through every cycle of a complete, independent set of cycles—a minimum cycle basis [7,8]—and analyse its statistics.

The minimum cycle basis is obtained with the standard algorithm [7,8], with each link weighted by its Euclidean length in order to compose the basis containing the geometrically shortest independent cycles. Because every cycle of the network is a combination of the basis cycles, the basis fluxes carry the complete gauge-invariant content of the hopping phases. While each basis cycle C is oriented counterclockwise, its flux $\Phi_C$ is defined as the argument of the product of the complex hoppings around the cycle, which reduces to the uniform plaquette flux of the standard Hofstadter lattice. To examine the relationship between the obtained flux and the cycle geometry, we record the enclosed area $A_C$ and the centroid distance $r_C$ of every cycle, normalized by the convex-hull area and the distance from its centre to the outermost node, respectively.

Figure S7 shows the resulting statistics for the optimized lattices. For the NN and NNN lattices, the designed gauge field is locked with distinct degrees: about 67% (NN; Fig. S7a) and 49% (NNN; Fig. S7b) of all cycles fall into a single flux bin of width $\pi/10$, with mean fluxes of $\Phi_C \approx -0.7\pi$ and $\Phi_C \approx -0.4\pi$, respectively. Regarding their spatial distributions, NN lattices exhibit highly uniform flux profiles, characterized by a narrow bandwidth of $\Phi_C$ over a broad range of $r_C$ (Fig. S7e), whereas NNN lattices exhibit weakly disordered flux distributions (Fig. S7f). Notably, the flux density $\Phi_C/A_C$ is nearly the same in the NN and NNN lattices, inheriting the spatially uniform effective magnetic field of the standard Hofstadter lattice.

In contrast, the optimized flux on the minimal cycles is broadly distributed in the 3NN and 4NN lattices (Figs. S7c,d,g,h). Although the fluxes here are not periodically staggered, such a configuration is reminiscent of the magnetic-field-free realization of the Haldane model, in which the net flux per unit cell vanishes while local fluxes do not [9]. In the present designs, however, the cancellation is statistical rather than exact within each unit cell, while the flux through the unit plaquettes of the underlying square lattice is as broadly distributed as that of the minimal cycles. This observation suggests that increasing the node degree, and thereby the number of crossing links, drives the topological gauge field from the uniform (NN) through the weakly disordered (NNN) Hofstadter type to the strongly disordered Haldane type (3NN and 4NN).

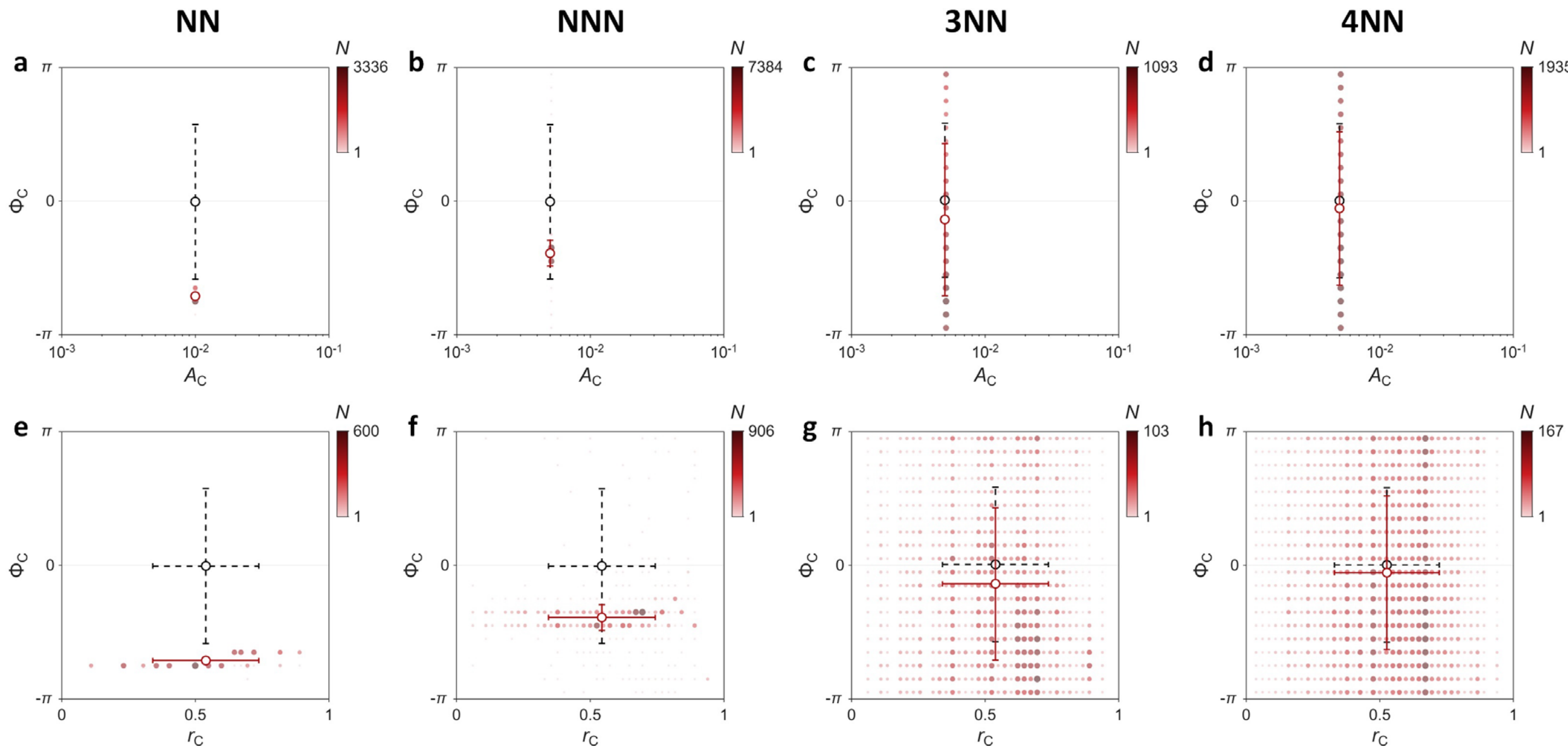

**Fig. S7. Holonomy statistics of the optimized lattices.** Cycle flux $\Phi_C$ of every cycle of the minimum cycle basis of the optimized lattices, pooled over 50 realizations, plotted against the normalized cycle area $A_C$ (**a-d**) and the normalized centroid radius $r_C$ (**e-h**), for the NN (**a,e**), NNN (**b,f**), 3NN (**c,g**), and 4NN (**d,h**) designs with $C_L^0 = 1$ and $\beta = 1$. Circle size and colour present the number $N$ of cycles per bin. Open circles with bars show the mean and one standard deviation of the optimized (red, solid) and initial random-phase (black, dashed) lattices. Cycles enclosing zero area are excluded. All other parameters are the same as those in Fig. 2.

**Supplementary Note S6. Degree-*k* network topology**

The networks studied in Figs. 1, 3, and 4 of the main text are generated on a one-parameter axis $\rho$ that connects a regular network to a random network (Fig. S8). We start from $N = 100$ nodes arranged as a $k$-regular ring. The ring is randomized by degree-preserving double-edge swaps: two links are chosen at random, and their end nodes are exchanged. The move is rejected if the four end nodes are not distinct or if a new link already exists (Fig. S8a). We quantify this randomization by the rewiring number per link, $\rho$.

Figure S8b shows three $k = 8$ networks at $\rho = 0$, 0.1, and 10. At $\rho = 0.1$, a few shortcuts are added to the ring, whereas almost all links have been rewired at $\rho = 10$. We note that node size and colour encode the betweenness centrality [10]—the fraction of shortest paths passing through each node—which highlights the hubs created by the rewiring.

To locate the small-world regime [11] of these degree-$k$ networks, we evaluate the small-worldness index $\omega$:

$$\omega = \frac{L_{\text{Random}}}{L} - \frac{C}{C_{\text{Regular}}}, \tag{S5}$$

where $L$ and $C$ are the mean shortest-path length and the clustering coefficient of the network, $L_{\text{Random}}$ is the path length of a random network, and $C_{\text{Regular}}$ is the clustering coefficient of the regular network. The regular and random networks are separated by the small-world regime near $\omega \approx 0$. Figure S8c shows $\omega$ over the $(k, \rho)$ grid, averaged over the four independent graph realizations of each cell. The result confirms that $\rho = 0$, 0.1, and 10 correspond to regular, small-world, and random networks, respectively.

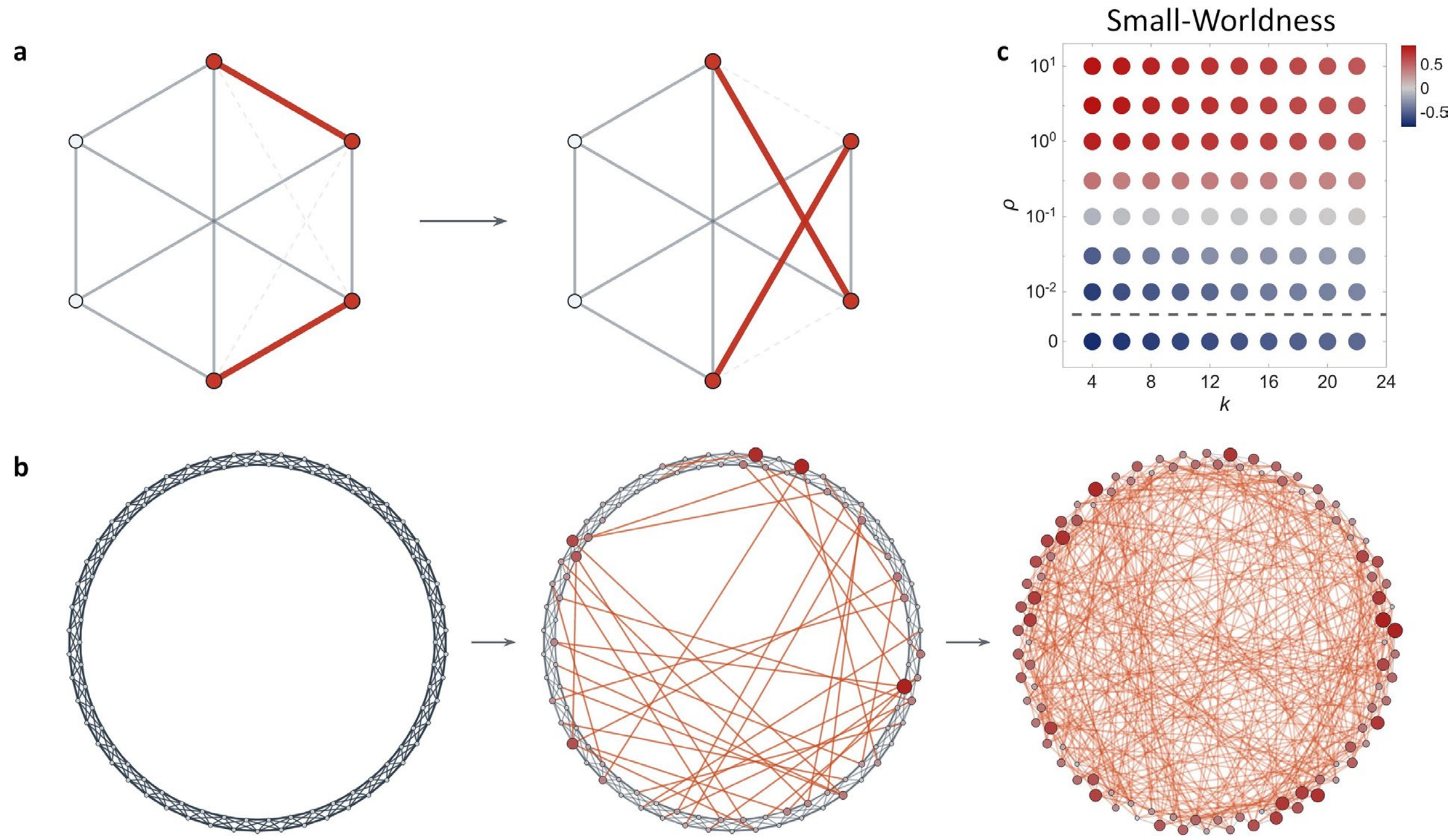

**Fig. S8. Regular-to-random network configuration.** **a**, A degree-preserving double-edge swap; removed (left) and introduced (right) links are shown in red. **b**, $k = 8$ networks ($N = 100$) after 0, 20, and 2000 swaps ($\rho = 0$, 0.1 and 10). Retained links are dark grey and swapped-in links are orange. Node size and colour scale with betweenness centrality. **c**, Small-worldness $\omega$ over ($k$, $\rho$), averaged over the four independent graphs of each cell. The dashed line separates the regular network ($\rho = 0$) from the rewired networks. All other parameters are the same as those in Figs. 3e-g.

**Supplementary Note S7. Network Hofstadter butterflies**

Figures S9 and S10 extend the high-degree Hofstadter butterfly of Fig. 2c in the main text to the network configurations of Fig. 3. Despite differences in detail, the $C_L$ maps are similar across the network topologies quantified by $\rho$ (Fig. S9): every network exhibits a plateau of nontrivial $C_L$ around the design point ($\chi = 1$ and $E = 1$).

The robustness of this plateau, however, depends strongly on the node degree $k$: a higher $k$ yields a larger local gap $\mu_L$ (Fig. S10), confirming that the node degree acts as the resource for a topologically nontrivial bulk with a stable local gap. The node degree $k$ also yields the spreading of eigenenergies under the unity link weight $t_{mn} = 1$.

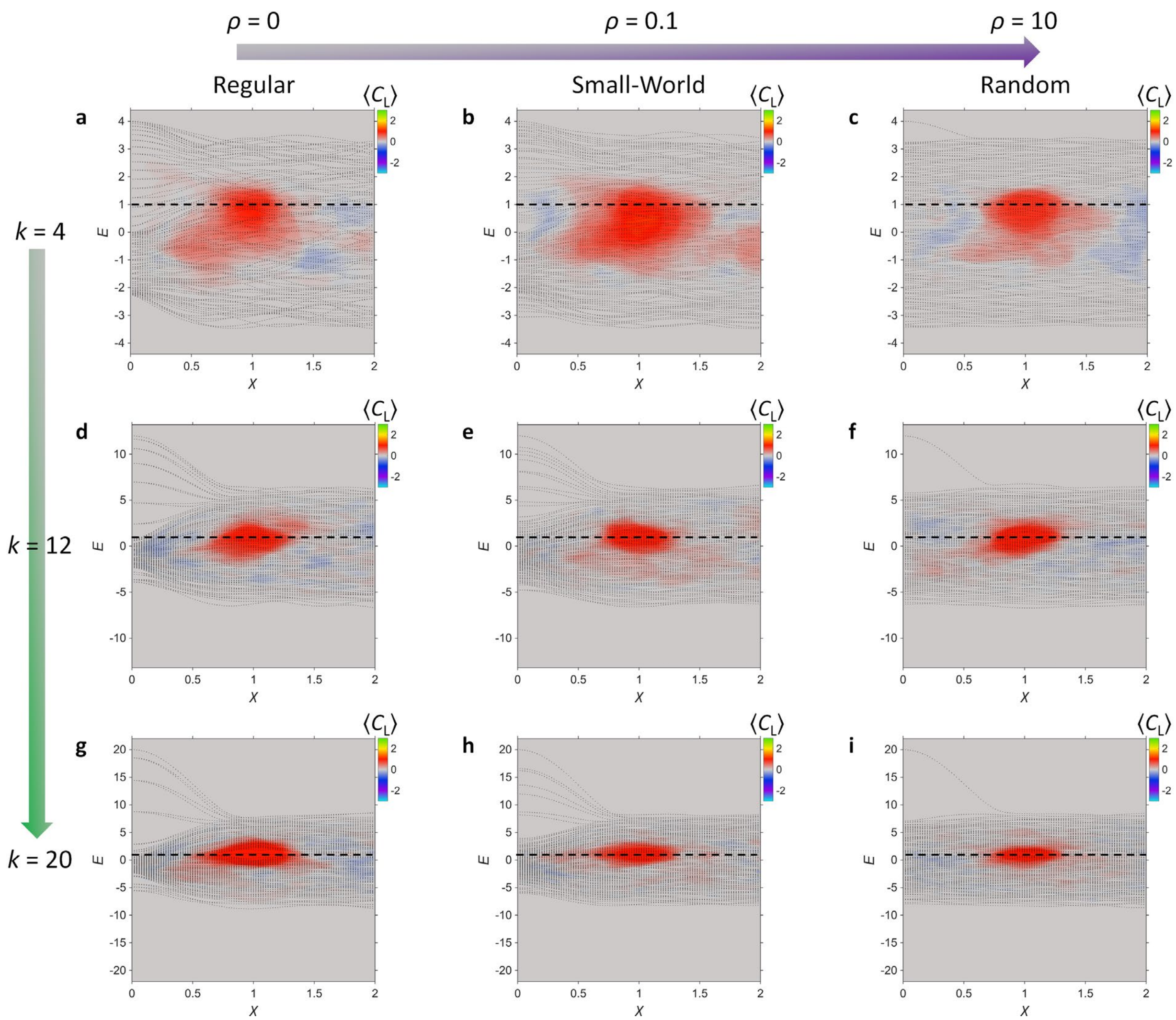

**Fig. S9. Network Hofstadter butterflies of the local Chern index.** Bulk-averaged $C_L$ as a function of the phase-rescaling factor $\chi$ and the energy $E$ of optimized networks with $k = 4$ (**a**-**c**), 12 (**d**-**f**), and 20 (**g**-**h**), which correspond to NN, 3NN, and 4NN in high-degree Hofstadter networks. Regular ($\rho = 0$; **a**,**d**,**g**), small-world ($\rho = 0.1$; **b**,**e**,**h**), and random ($\rho = 10$; **c**,**f**,**i**) networks are examined. Points denote the eigenenergies; dashed lines mark $E = 1$. All other parameters are the same as those in Fig. 3.

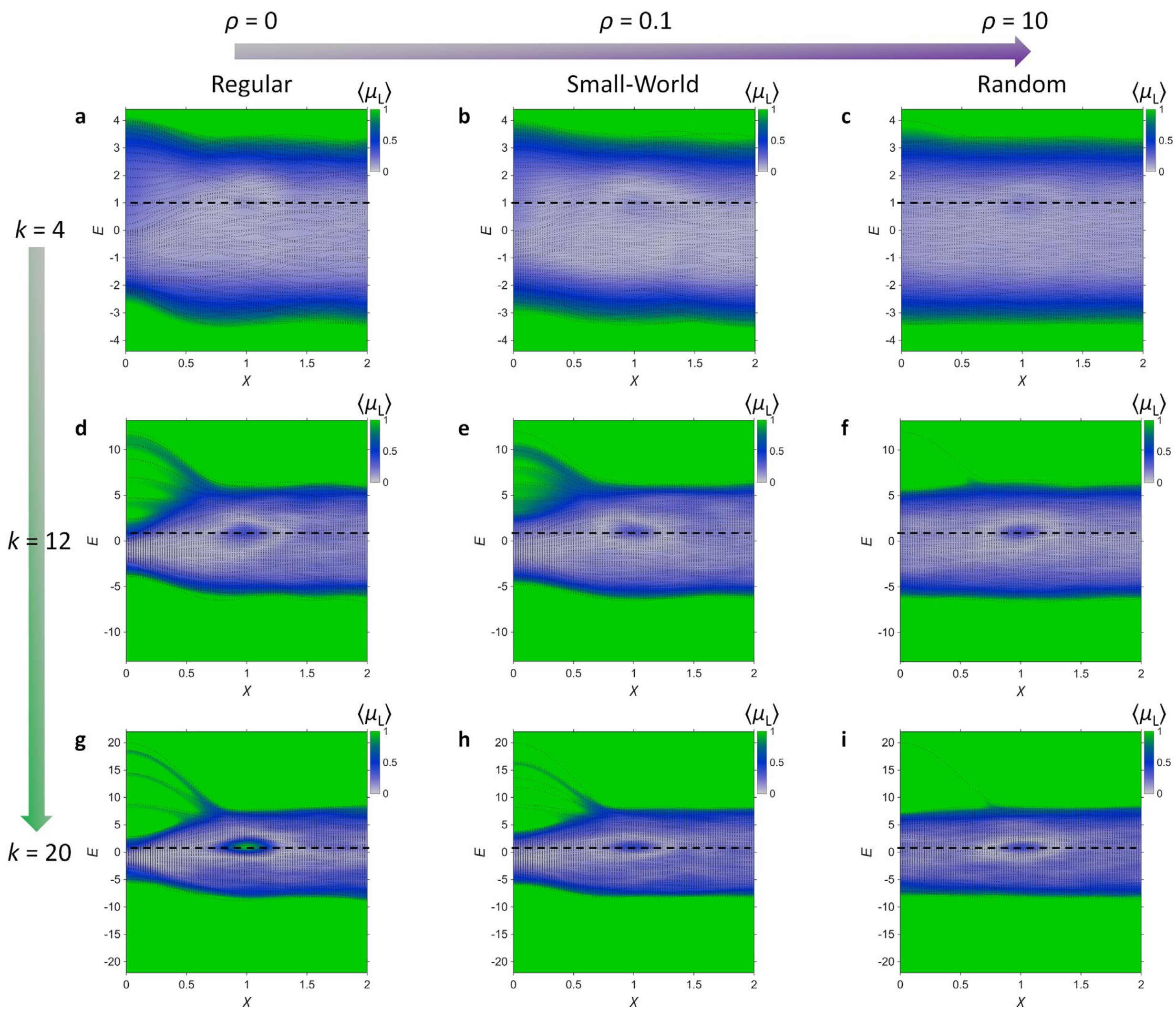

**Fig. S10. Network Hofstadter butterflies of the local gap.** Bulk-averaged $\mu_{\mathrm{L}}$ for the same realizations, axes, and conventions as in Fig. S9.

**Supplementary Note S8. Network optimization curves**

Figure 1c of the main text shows the optimization history for a single small-world network realization. For completeness, Fig. S11 presents the corresponding ensemble-averaged optimization curves for small-world networks: $k = 4$ and $k = 20$ at $\rho = 0.1$ and $\beta = 10$. During the GDS, our optimization policy successfully achieves nontrivial topology, $C_\beta \approx C_L^0 = 1$, for both node degrees. However, enhancement of the local gap occurs only at the higher node degree, $k = 20$. These optimization curves are consistent with the main results presented in Fig. 3 of the main text.

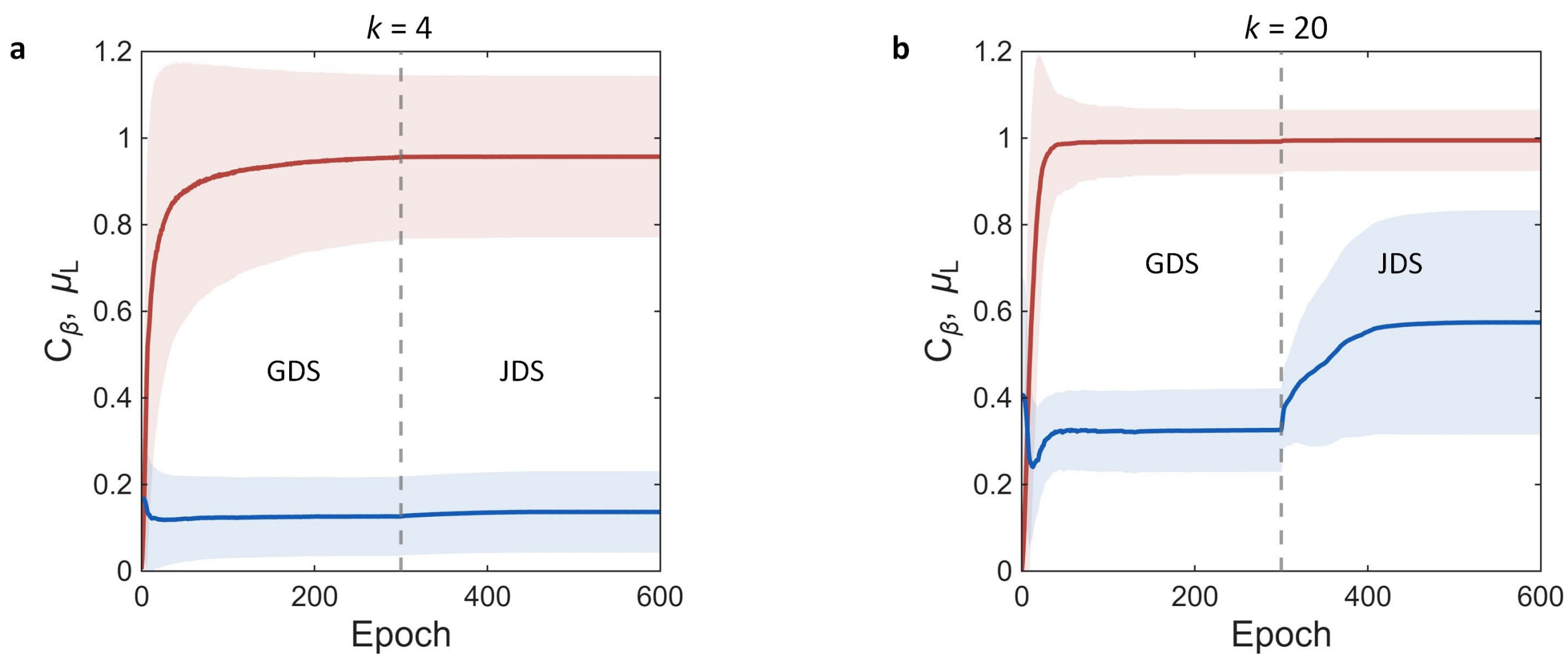

**Fig. S11. Ensemble optimization curves for network designs.** Mean (lines) and one standard deviation (shaded) of the bulk $C_\beta$ (red) and $\mu_L$ (blue), pooled over the 60 bulk nodes of 40 realizations at $\rho = 0.1$ and $\beta = 10$, for $k = 4$ (**a**) and $k = 20$ (**b**). Epochs count accepted updates; the dashed line marks the transition from the GDS stage (300 updates) to the JDS stage (300 updates). All other parameters are the same as those in Fig. 3.

**Supplementary Note S9. Localization and phase frustration**

As additional metrics for analysing the network topological insulators, we introduce two quantities: the inverse participation ratio (IPR) and the degree of phase frustration. First, the IPR, which increases for more localized eigenstates, is defined as

$$\mathrm{IPR} = N\sum_{r=1}^{N}\left|\psi_r\right|^4, \tag{S6}$$

where $\psi_r$ is the amplitude at node $r$ of the normalized eigenstate whose eigenenergy is closest to $E = 1$. Second, using the magnetic adjacency matrix $A = H$ and the degree matrix $D = kI$ of our degree-$k$ networks, we introduce the normalized magnetic Laplacian [12] and its lowest eigenvalue, as follows:

$$L_n = D^{-1/2}\left(D - A\right)D^{-1/2}, \qquad \lambda_{\mathrm{G}} = \min \mathrm{Spec}\left(L_n\right). \tag{S7}$$

Although $\lambda_{\mathrm{G}}$ bounds, through the magnetic Cheeger inequality [13,14], the minimal cost of isolating a large phase-coherent subgraph from the network, in our case its change during the optimization measures only the degree of phase frustration, because the network connectivity is unchanged by the optimization.

Figure S12a shows the IPR over the ($k$, $\rho$) grid. Notably, the eigenstates of the regular networks ($\rho = 0$) are the most localized, with the largest IPR. A comparison with the edge confinement in Fig. 3g of the main text shows that this localization has a different origin: topological edge states appear only at higher node degrees, whereas the strong localization of the regular networks at low $k$ is not accompanied by edge confinement.

More interesting results are observed in the optimization-induced change of $\lambda_{\mathrm{G}}$ in Fig. S12b. The separation cost—corresponding to the degree of phase frustration—increases at higher node degrees. Because the phase frustration originates from nonzero fluxes through the network cycles, this result also supports the emergence of nontrivial topology in high-*k* regimes.

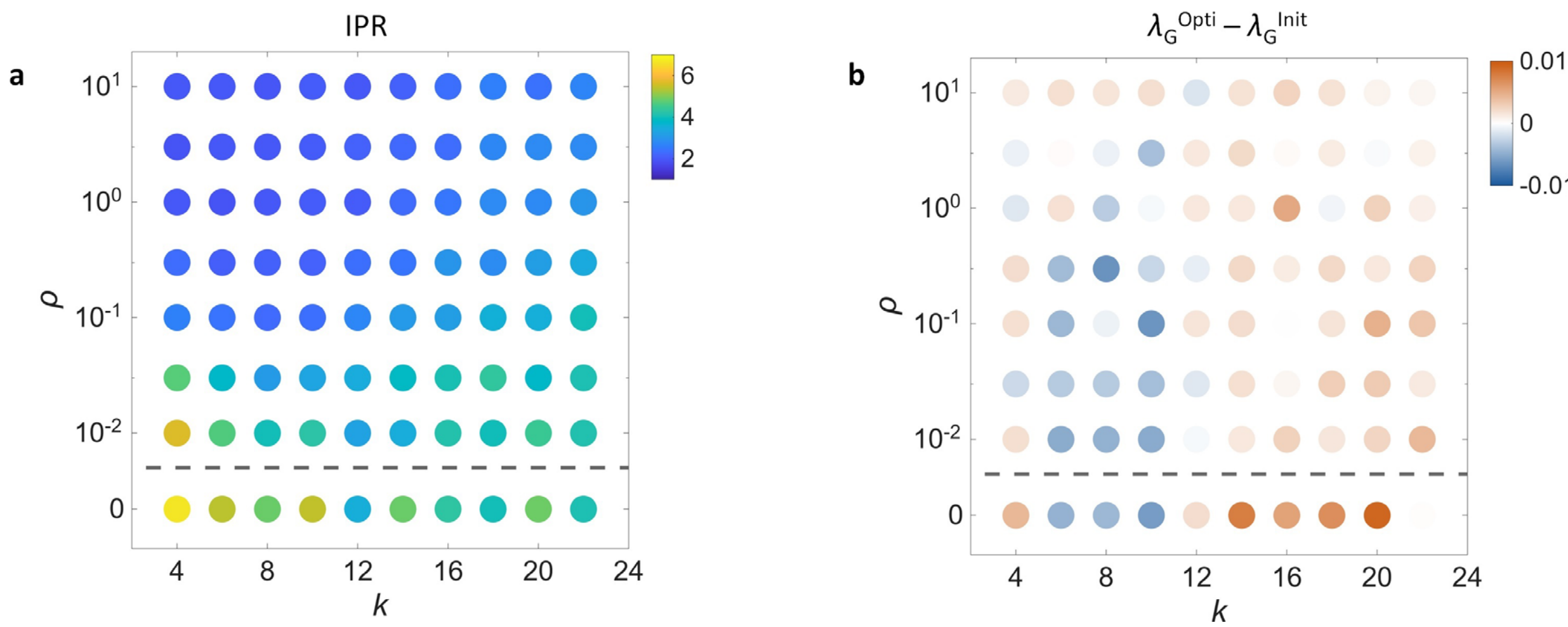

**Fig. S12. Localization and phase frustration across the design space. a**, IPR. **b**, Change of $\lambda^G$ of the normalized magnetic Laplacian between the optimized and the initial networks. Each disc is the mean over 40 realizations per ($k$, $\rho$) cell. All other parameters are the same as those in Fig. 3e–g.

**Supplementary Note S10. Spectral responses**

Figures S13 and S14 show the spectrally resolved edge confinement and localization, respectively, for the initial and optimized networks. Both metrics demonstrate the emergence of topological edge states at the design energy $E = 1$ for higher node degrees $k$. In contrast, the IPR of the initial networks is already large near the band edges of the spectrum and for $k = 4$, showing that these localized states do not originate from nontrivial topology.

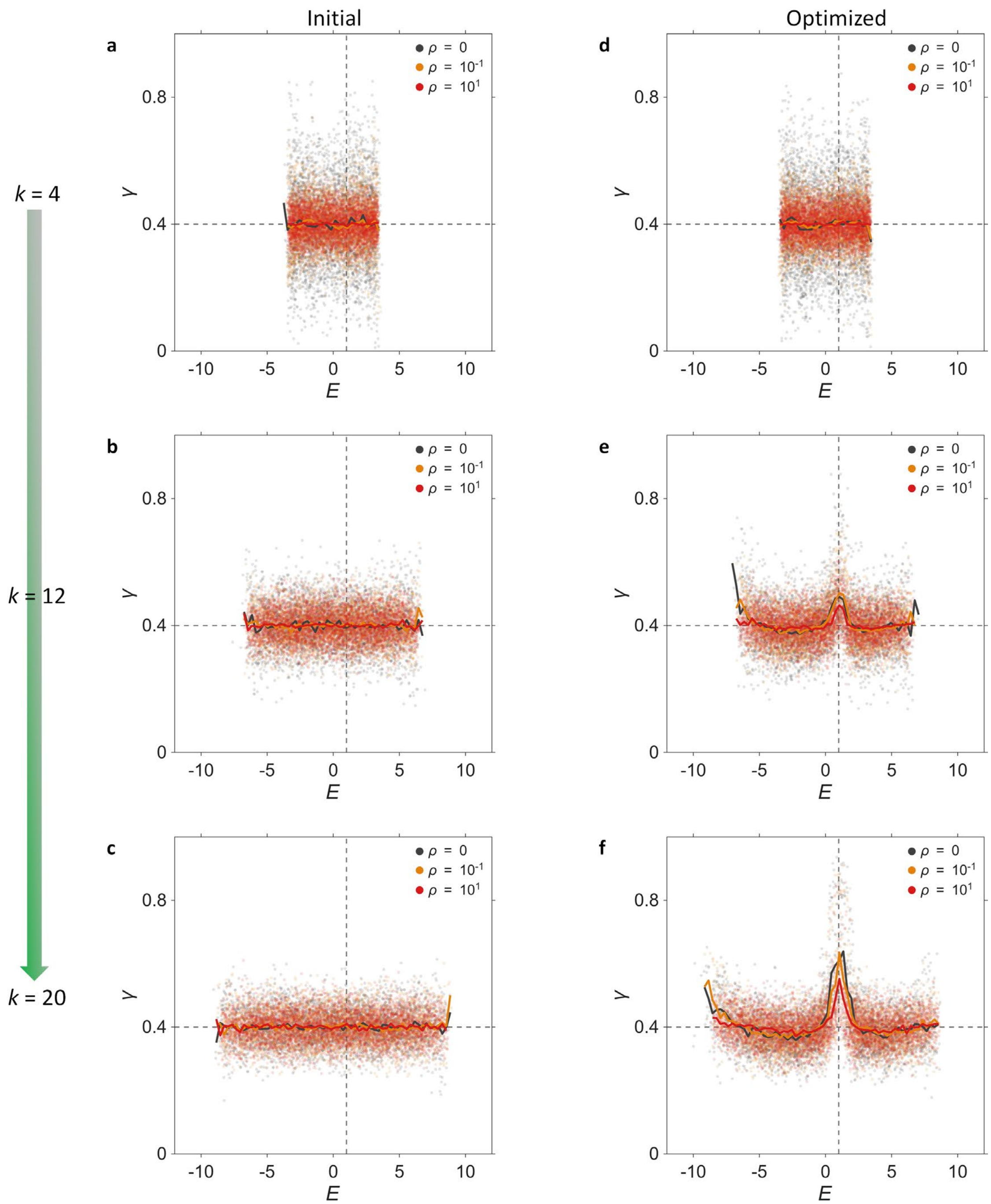

**Fig. S13. Spectrally resolved edge confinement.** Edge confinement $\gamma$ of every eigenstate as a function of its eigenenergy $E$, before (**a-c**) and after (**d-f**) optimization, for $k = 4$ (**a,d**), 12 (**b,e**), and 20 (**c,f**), with $\rho = 0$ (grey), 0.1 (orange), and 10 (red). Each point cloud pools all 100 eigenstates from each of 40 realizations; solid lines indicate ensemble means. Dashed lines mark $E = 1$ and the uniform-distribution value $\gamma = 0.4$. All other parameters are the same as those in Fig. 3.

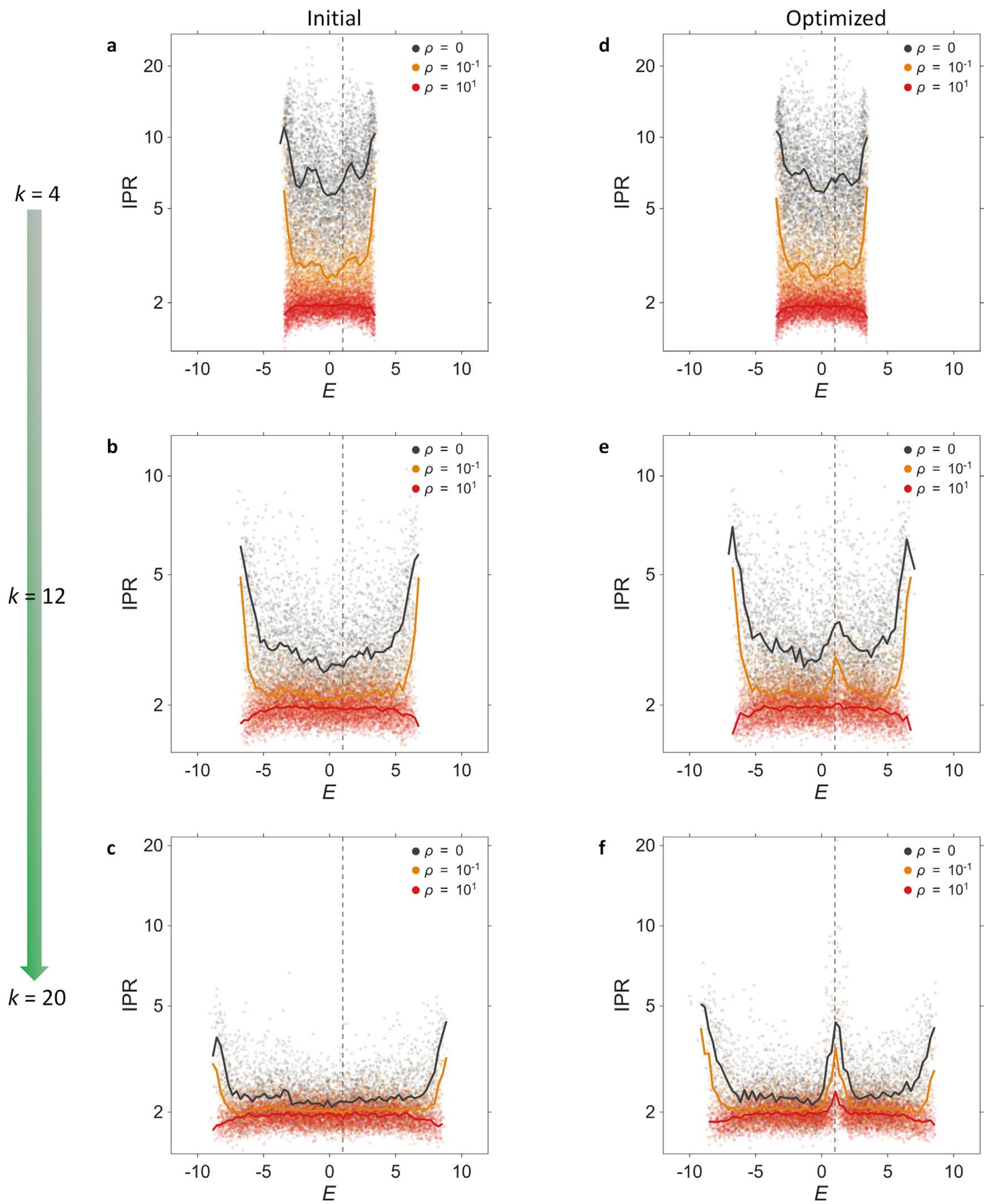

**Fig. S14. Spectrally resolved localization.** IPR of every eigenstate as a function of its eigenenergy *E*, before (**a-c**) and after (**d-f**) optimization, for the same networks, realizations, and colour coding as in Fig. S13. The IPR axis is logarithmic.

## Supplementary Note S11. Flux evaluations in network topological insulators

As in Note S5, we perform a holonomy analysis of the $k$ = 20 networks before and after their optimization into network topological insulators. In all three regimes—regular, small-world, and random—the fluxes through the minimal cycles are distributed nearly uniformly over $(-\pi, \pi]$ both before and after optimization. As in the 3NN and 4NN lattices, which realize Haldane-type Chern-insulator phases, the flux $\Phi_C$ remains very small, largely independent of network topology.

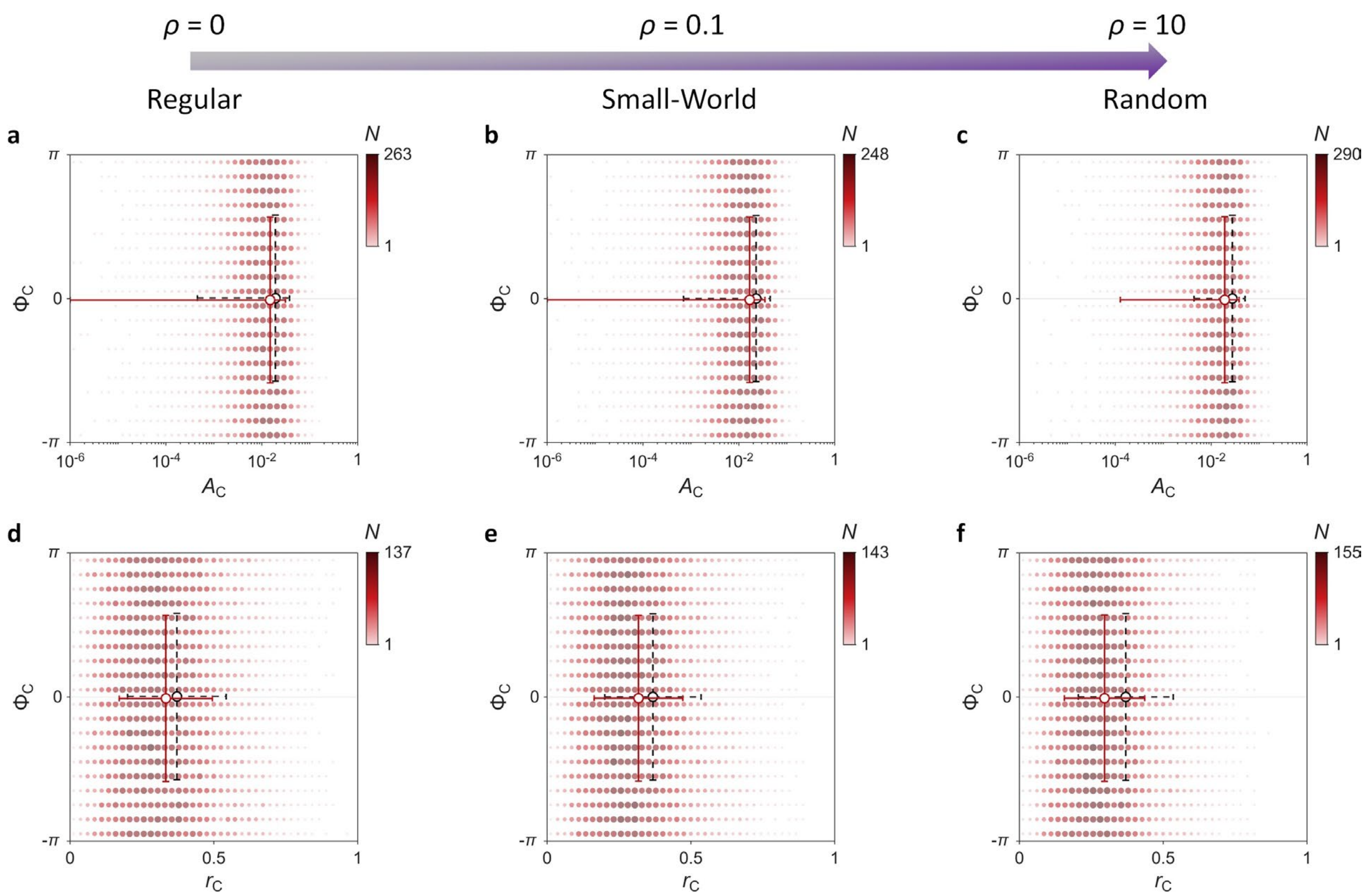

**Fig. S15. Holonomy statistics of the optimized networks.** Cycle flux $\Phi_C$ of every cycle of the minimum cycle basis of the optimized $k = 20$ networks, pooled over 40 realizations, plotted against the normalized cycle area $A_C$ (**a-c**) and the normalized centroid radius $r_C$ (**d-f**) for $\rho = 0$ (**a,d**), 0.1 (**b,e**), and 10 (**c,f**). Conventions are as in Fig. S7. All other parameters are the same as those in Fig. 3.

**Supplementary Note S12. Weakly coupled excitation**

Although the strongly coupled multiport geometry of Fig. 4 in the main text is the suitable configuration for measuring chiral transport, this strong coupling at the same time perturbs the eigenstates, hindering the visualization of unperturbed eigenstate evolutions. Therefore, we analyse the weakly coupled, single-port excitation of the small-world topological network of Fig. 4 ($k$ = 20, ρ = 0.1) at $E$ = 1 (Fig. S16). With a single port, the wave is fully reflected with the scattering coefficient $|S_{11}|$ = 1 with the nearly unperturbed profile of the eigenstate near $E$ = 1. The result shows the strong confinement of the excited-state intensity $|\psi|^2$ at the network edge nodes, which belong to the $C_L$ = 0 region. The corresponding Supplementary Movie S3 animates Re[$\psi$exp($-iv$)] for $v$ from 0 to 2π, visualizing the chiral phase evolution along the network edge.

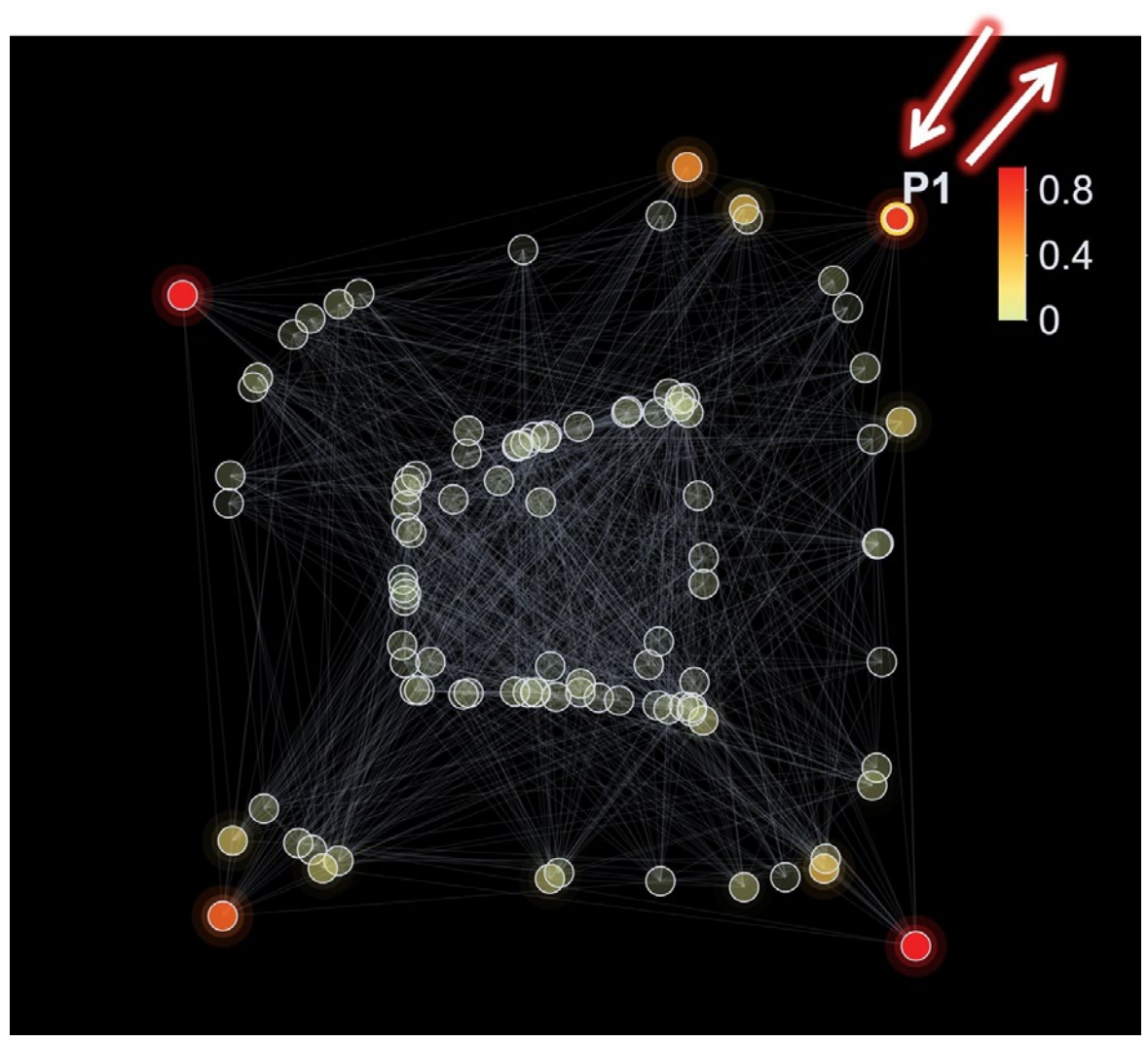

**Fig. S16. Single-port excitation of the small-world topological insulator.** Steady-state intensity $|\psi|^2$ when the device of Fig. 4 is driven through a single port $P_1$ at $E$ = 1 with the weak external coupling $t_{ext}$ = 0.25 (< $t_{mn}$ = 1). The wave is fully reflected, while the excitation is confined on the edge state. Node colour and opacity scale with $|\psi|^2$. Arrows indicate the incident and reflected waves. All other parameters are the same as those in Fig. 4.

## Supplementary Note S13. Extended analysis of multiport transport in network routers

Figure S17 shows the transmission matrices at $E = 1$. For the realization of each regime with the largest chirality $|\eta|$ at $E = 1$ among the 40 realizations (Figs. S17a-c), the small-world device acts as a nearly ideal circulator: the transmission to the next counterclockwise port is close to unity and all other elements are close to zero. The regular device shows the same routing pattern with leakage between ports 1 to 3, and the random device shows a weaker element with residual reflection. The ensemble averages (Figs. S17d-f) retain the counterclockwise routing pattern, with increasingly populated off-route elements from the regular to the random regime.

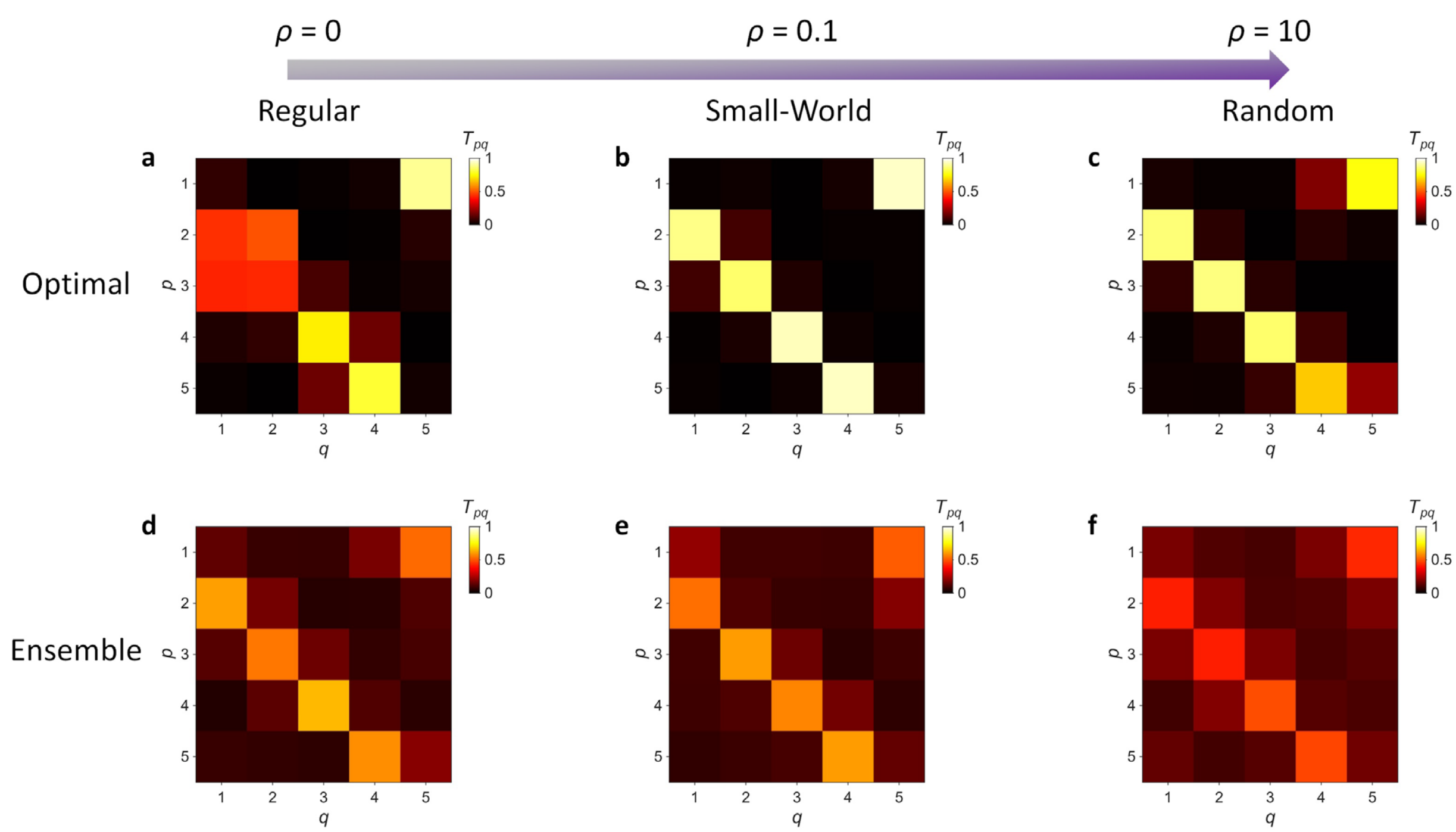

**Fig. S17. Port-to-port transmission of network routers.** Transmission matrices $T_{pq} = |S_{pq}|^2$ at $E = 1$ for five-port routers built on optimized $k = 20$ networks. **a-c**, The realization with the largest $|\eta|$ at $E = 1$ in each regime, at its individually selected external coupling $t_{\text{ext}} = 2.375$ (**a**), 2.493 (**b**), and 2.965 (**c**). **d-f**, Averages over the 40 realizations at each regime at $t_{\text{ext}} = 2.375$ (**d**), 2.611 (**e**), and 2.729 (**f**). Network topology: regular ($\rho = 0$; **a,d**), small-world ($\rho = 0.1$; **b,e**), and random ($\rho = 10$; **c,f**) networks. All other parameters are the same as those in Fig. 4.

Figure S18 shows the corresponding spectra $T_{\rm CCW}(E)$ and $T_{\rm CW}(E)$. The optimal small-world device achieves the best performance: $T_{\rm CCW} \approx 0.9$ with $T_{\rm CW} \approx 0$ over a broad energy window $E \approx$ 0.4 to 1.9 around the design energy $E = 1$. The ensemble-averaged $T_{\rm CCW}$ peaks and $T_{\rm CCW} > T_{\rm CW}$ near $E = 1$ in all regimes of network topology, showing that the chiral routing is a robust property of the ensemble across the entire regular-to-random axis. Far from the design energy, the ensemble-averaged $T_{\rm CCW}$ and $T_{\rm CW}$ approach each other, consistent with a nonreciprocity confined to the energy window of the designed gap.

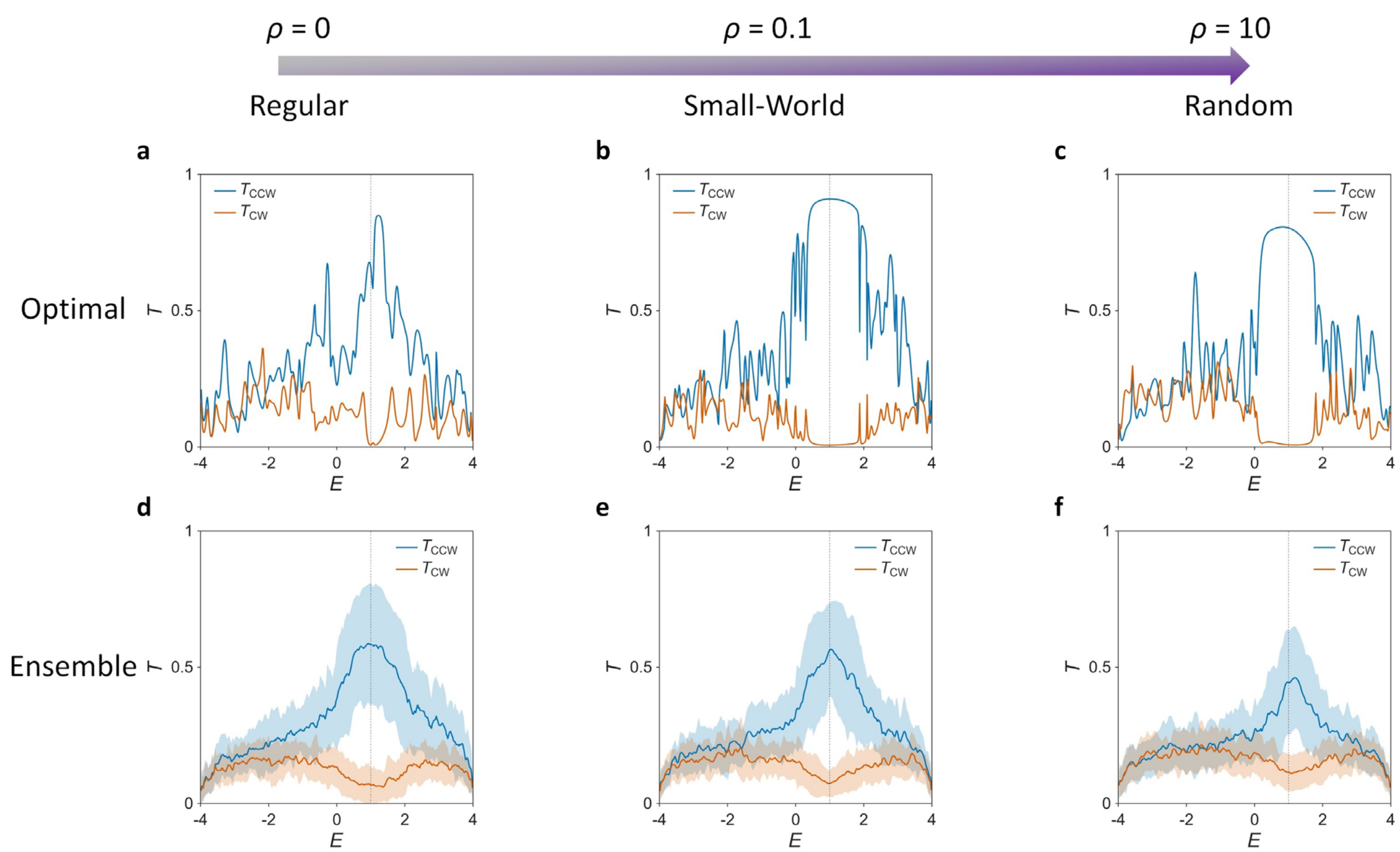

**Fig. S18. Chiral transport spectra of network routers.** Counterclockwise ($T_{\rm CCW}$, blue) and clockwise ($T_{\rm CW}$, orange) transmission between adjacent ports, averaged over input ports, versus energy $E$ for the same single realizations (**a-c**, optimal cases) and ensembles (**d-f**; lines, mean; shading, one standard deviation over the 40 realizations) as in Fig. S17. Dashed lines mark $E = 1$. All other parameters are the same as those in Fig. S17.

**Supplementary Note S14. Robustness of regular and random topological insulators**

Figure S19 examines the disorder immunity of the regular and random topological insulators with the realizations of optimal $\eta$. In sharp contrast to the small-world one, these architectures are rigorously (regular) or statistically (random) homogeneous in their node centralities. This microstructural difference is directly reflected in the robustness of the chirality. In the regular network, whose node and link centralities are completely degenerate, the two groups are statistically indistinguishable under every perturbation. In the random network, the differentiation is substantially suppressed for the onsite and hopping-phase perturbations—by factors of four and three relative to the small-world network—whereas it persists under the hopping-amplitude perturbation.

Another critical feature of the network topological insulators is the superior robustness of the chirality in the small-world (Figs. 4e-g) and random (Figs. S19d-f) networks compared with the regular network (Figs. S19a-c). Considering the classification of the small-world regime in Eq. (S5), this robustness follows the reduction of the characteristic path length from the regular network to the small-world and random networks, which shortens the edge path through shortcut connectivity. Despite the critical role of the path length, the ordering in robustness is not set by the path length alone. the random network has the shortest paths yet is less robust than the small-world one, suggesting that the robustness is maximal where the strong clustering and short path length coexist.

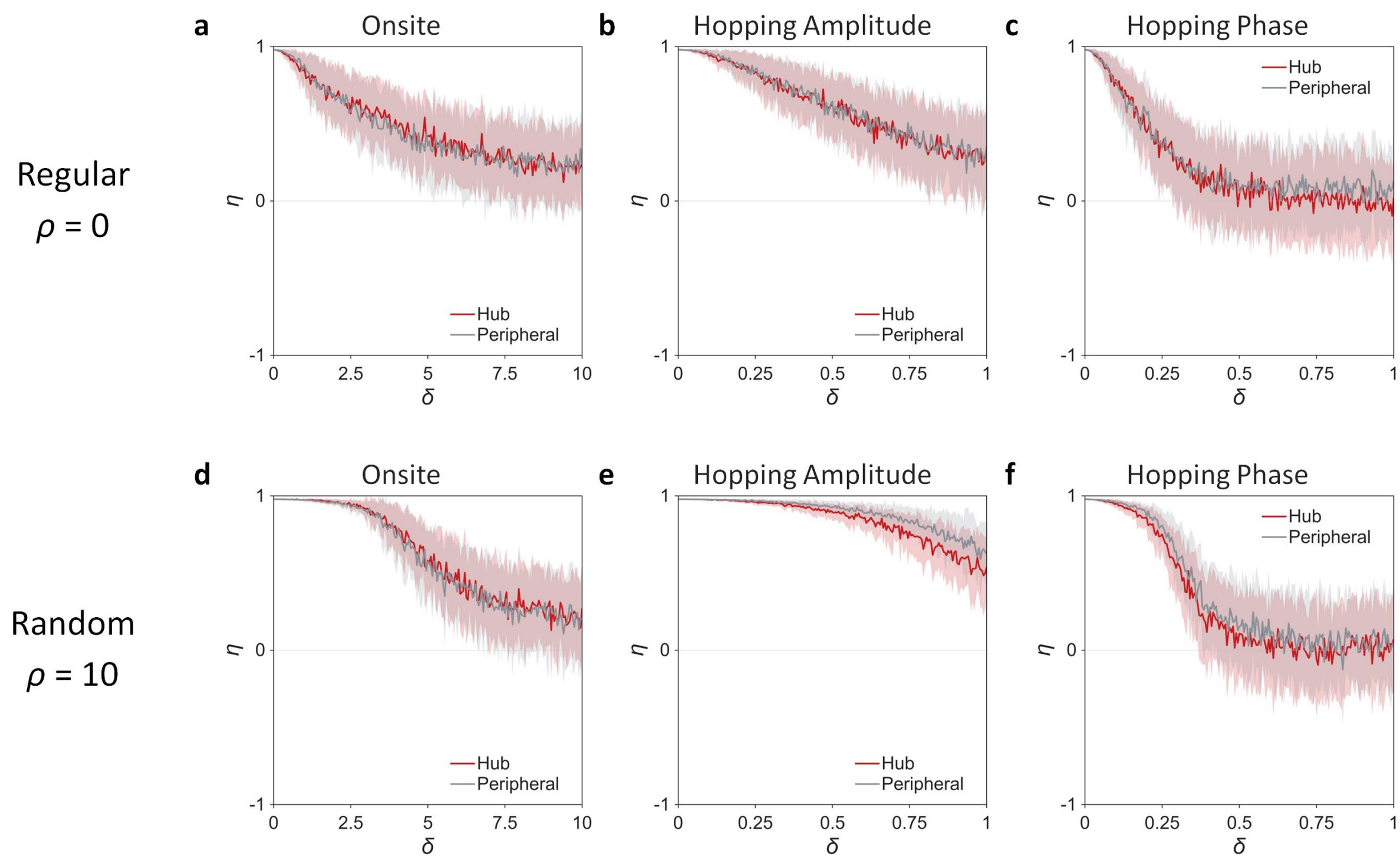

**Fig. S19. Regular and random topological insulators are centrality homogeneous.** $\eta(E = 1)$ of the regular (**a-c**) and random (**d-f**) network topological insulators under onsite (**a,d**), hopping-amplitude (**b,e**), and hopping-phase (**c,f**) perturbations of strength $\delta$ applied to the most central half (hubs, red) or the least central half (peripheral nodes, grey) of the network, at $t_{\text{ext}} = 2.375$ (**a-c**) or 2.729 (**d-f**). All other parameters and conventions are the same as those in Fig. 4.

## Supplementary References

preprint arXiv:2607.04736 (2026).